\documentclass[twocolumn,showpacs,preprintnumbers,amsmath,prl,amssymb,superscriptaddress]{revtex4}

\pdfoutput=1

\usepackage{graphicx}
\usepackage{dcolumn}
\usepackage{bm}

\usepackage{enumitem}
\setenumerate{itemindent=2em,leftmargin=0pt,itemsep=0pt,partopsep=0pt}
\setitemize[1]{itemindent=2em,leftmargin=2pt,itemsep=0pt,partopsep=0pt,topsep=0pt}

\def\effbs{\varepsilon _{\rm BS}}
\def\lmbdbs{\lambda _{\rm BS}}

\begin{document}


\title{A Search of Low-Mass WIMPs with p-type Point Contact Germanium Detector in the CDEX-1 Experiment
}


\affiliation{Key Laboratory of Particle and Radiation Imaging (Ministry of Education) and Department of Engineering Physics, Tsinghua University, Beijing 100084}
\affiliation{College of Physical Science and Technology, Sichuan University, Chengdu 610064}
\affiliation{Department of Nuclear Physics, China Institute of Atomic Energy, Beijing 102413}
\affiliation{School of Physics, Nankai University, Tianjin 300071}
\affiliation{NUCTECH Company, Beijing 10084}
\affiliation{YaLong River Hydropower Development Company, Chengdu 610051}
\affiliation{Institute of Physics, Academia Sinica, Taipei 11529}
\affiliation{Department of Physics, Banaras Hindu University, Varanasi 221005}
\author{W. Zhao}
\affiliation{Key Laboratory of Particle and Radiation Imaging (Ministry of Education) and Department of Engineering Physics, Tsinghua University, Beijing 100084}
\author{Q. Yue}\altaffiliation [Corresponding author: ]{yueq@mail.tsinghua.edu.cn}
\affiliation{Key Laboratory of Particle and Radiation Imaging (Ministry of Education) and Department of Engineering Physics, Tsinghua University, Beijing 100084}
\author{K.J.~Kang}
\affiliation{Key Laboratory of Particle and Radiation Imaging (Ministry of Education) and Department of Engineering Physics, Tsinghua University, Beijing 100084}
\author{J.P.~Cheng}
\affiliation{Key Laboratory of Particle and Radiation Imaging (Ministry of Education) and Department of Engineering Physics, Tsinghua University, Beijing 100084}
\author{Y.J.~Li}
\affiliation{Key Laboratory of Particle and Radiation Imaging (Ministry of Education) and Department of Engineering Physics, Tsinghua University, Beijing 100084}
\author{H.T.~Wong}
\altaffiliation{Participating as a member of TEXONO Collaboration}
\affiliation{Institute of Physics, Academia Sinica, Taipei 11529}
\author{S.T.~Lin}
\altaffiliation{Corresponding author: linst@phys.sinica.edu.tw}
\affiliation{College of Physical Science and Technology, Sichuan University, Chengdu 610064}
\affiliation{Institute of Physics, Academia Sinica, Taipei 11529}
\author{J.P.~Chang}
\affiliation{NUCTECH Company, Beijing 10084}
\author{J.H.~Chen}
\altaffiliation{Participating as a member of TEXONO Collaboration}
\affiliation{Institute of Physics, Academia Sinica, Taipei 11529}
\author{Q.H.~Chen}
\affiliation{Key Laboratory of Particle and Radiation Imaging (Ministry of Education) and Department of Engineering Physics, Tsinghua University, Beijing 100084}
\author{Y.H.~Chen}
\affiliation{YaLong River Hydropower Development Company, Chengdu 610051}
\author{Z.~Deng}
\affiliation{Key Laboratory of Particle and Radiation Imaging (Ministry of Education) and Department of Engineering Physics, Tsinghua University, Beijing 100084}
\author{Q.~Du}
\affiliation{School of Physics, Nankai University, Tianjin 300071}
\author{H.~Gong}
\affiliation{Key Laboratory of Particle and Radiation Imaging (Ministry of Education) and Department of Engineering Physics, Tsinghua University, Beijing 100084}
\author{X.Q.~Hao}
\affiliation{Key Laboratory of Particle and Radiation Imaging (Ministry of Education) and Department of Engineering Physics, Tsinghua University, Beijing 100084}
\author{H.J.~He}
\affiliation{Key Laboratory of Particle and Radiation Imaging (Ministry of Education) and Department of Engineering Physics, Tsinghua University, Beijing 100084}
\author{Q.J.~He}
\affiliation{Key Laboratory of Particle and Radiation Imaging (Ministry of Education) and Department of Engineering Physics, Tsinghua University, Beijing 100084}
\author{H.X.~Huang}
\affiliation{Department of Nuclear Physics, China Institute of Atomic Energy, Beijing 102413}
\author{T.R.~Huang}
\altaffiliation{Participating as a member of TEXONO Collaboration}
\affiliation{Institute of Physics, Academia Sinica, Taipei 11529}
\author{H.~Jiang}
\affiliation{Key Laboratory of Particle and Radiation Imaging (Ministry of Education) and Department of Engineering Physics, Tsinghua University, Beijing 100084}
\author{H.B.~Li}
\altaffiliation{Participating as a member of TEXONO Collaboration}
\affiliation{Institute of Physics, Academia Sinica, Taipei 11529}
\author{J.~Li}
\affiliation{Key Laboratory of Particle and Radiation Imaging (Ministry of Education) and Department of Engineering Physics, Tsinghua University, Beijing 100084}
\author{J.~Li}
\affiliation{NUCTECH Company, Beijing 10084}
\author{J.M.~Li}
\affiliation{Key Laboratory of Particle and Radiation Imaging (Ministry of Education) and Department of Engineering Physics, Tsinghua University, Beijing 100084}
\author{X.~Li}
\affiliation{Department of Nuclear Physics, China Institute of Atomic Energy, Beijing 102413}
\author{X.Y.~Li}
\affiliation{School of Physics, Nankai University, Tianjin 300071}
\author{Y.L.~Li}
\affiliation{Key Laboratory of Particle and Radiation Imaging (Ministry of Education) and Department of Engineering Physics, Tsinghua University, Beijing 100084}
\author{F.K.~Lin}
\altaffiliation{Participating as a member of TEXONO Collaboration}
\affiliation{Institute of Physics, Academia Sinica, Taipei 11529}
\author{S.K.~Liu}
\affiliation{College of Physical Science and Technology, Sichuan University, Chengdu 610064}
\author{L.C.~L\"{u}}
\affiliation{Key Laboratory of Particle and Radiation Imaging (Ministry of Education) and Department of Engineering Physics, Tsinghua University, Beijing 100084}
\author{H.~Ma}
\affiliation{Key Laboratory of Particle and Radiation Imaging (Ministry of Education) and Department of Engineering Physics, Tsinghua University, Beijing 100084}
\author{J.L.~Ma}
\affiliation{Key Laboratory of Particle and Radiation Imaging (Ministry of Education) and Department of Engineering Physics, Tsinghua University, Beijing 100084}
\author{S.J.~Mao}
\affiliation{NUCTECH Company, Beijing 10084}
\author{J.Q.~Qin}
\affiliation{Key Laboratory of Particle and Radiation Imaging (Ministry of Education) and Department of Engineering Physics, Tsinghua University, Beijing 100084}
\author{J.~Ren}
\affiliation{Department of Nuclear Physics, China Institute of Atomic Energy, Beijing 102413}
\author{J.~Ren}
\affiliation{Key Laboratory of Particle and Radiation Imaging (Ministry of Education) and Department of Engineering Physics, Tsinghua University, Beijing 100084}
\author{X.C.~Ruan}
\affiliation{Department of Nuclear Physics, China Institute of Atomic Energy, Beijing 102413}
\author{V.~Sharma}
\affiliation{Institute of Physics, Academia Sinica, Taipei 11529}
\affiliation{Department of Physics, Banaras Hindu University, Varanasi 221005}
\author{M.B.~Shen}
\affiliation{YaLong River Hydropower Development Company, Chengdu 610051}
\author{L.~Singh}
\altaffiliation{Participating as a member of TEXONO Collaboration}
\affiliation{Institute of Physics, Academia Sinica, Taipei 11529}
\affiliation{Department of Physics, Banaras Hindu University, Varanasi 221005}
\author{M.K.~Singh}
\altaffiliation{Participating as a member of TEXONO Collaboration}
\affiliation{Institute of Physics, Academia Sinica, Taipei 11529}
\author{A.K.~Soma}
\altaffiliation{Participating as a member of TEXONO Collaboration}
\affiliation{Institute of Physics, Academia Sinica, Taipei 11529}
\author{J.~Su}
\affiliation{Key Laboratory of Particle and Radiation Imaging (Ministry of Education) and Department of Engineering Physics, Tsinghua University, Beijing 100084}
\author{C.J.~Tang}
\affiliation{College of Physical Science and Technology, Sichuan University, Chengdu 610064}
\author{J.M.~Wang}
\affiliation{YaLong River Hydropower Development Company, Chengdu 610051}
\author{L.~Wang}
\affiliation{Key Laboratory of Particle and Radiation Imaging (Ministry of Education) and Department of Engineering Physics, Tsinghua University, Beijing 100084}
\author{Q.~Wang}
\affiliation{Key Laboratory of Particle and Radiation Imaging (Ministry of Education) and Department of Engineering Physics, Tsinghua University, Beijing 100084}
\author{S.Y.~Wu}
\affiliation{YaLong River Hydropower Development Company, Chengdu 610051}
\author{Y.C.~Wu}
\affiliation{NUCTECH Company, Beijing 10084}
\author{Z.Z.~Xianyu}
\affiliation{Key Laboratory of Particle and Radiation Imaging (Ministry of Education) and Department of Engineering Physics, Tsinghua University, Beijing 100084}
\author{R.Q.~Xiao}
\affiliation{Key Laboratory of Particle and Radiation Imaging (Ministry of Education) and Department of Engineering Physics, Tsinghua University, Beijing 100084}
\author{H.Y.~Xing}
\affiliation{College of Physical Science and Technology, Sichuan University, Chengdu 610064}
\author{F.Z.~Xu}
\affiliation{Key Laboratory of Particle and Radiation Imaging (Ministry of Education) and Department of Engineering Physics, Tsinghua University, Beijing 100084}
\author{Y.~Xu}
\affiliation{School of Physics, Nankai University, Tianjin 300071}
\author{X.J.~Xu}
\affiliation{Key Laboratory of Particle and Radiation Imaging (Ministry of Education) and Department of Engineering Physics, Tsinghua University, Beijing 100084}
\author{T.~Xue}
\affiliation{Key Laboratory of Particle and Radiation Imaging (Ministry of Education) and Department of Engineering Physics, Tsinghua University, Beijing 100084}
\author{L.T.~Yang}
\affiliation{Key Laboratory of Particle and Radiation Imaging (Ministry of Education) and Department of Engineering Physics, Tsinghua University, Beijing 100084}
\author{S.W.~Yang}
\altaffiliation{Participating as a member of TEXONO Collaboration}
\affiliation{Institute of Physics, Academia Sinica, Taipei 11529}
\author{N.~Yi}
\affiliation{Key Laboratory of Particle and Radiation Imaging (Ministry of Education) and Department of Engineering Physics, Tsinghua University, Beijing 100084}
\author{C.X.~Yu}
\affiliation{School of Physics, Nankai University, Tianjin 300071}
\author{H.~Yu}
\affiliation{Key Laboratory of Particle and Radiation Imaging (Ministry of Education) and Department of Engineering Physics, Tsinghua University, Beijing 100084}
\author{X.Z.~Yu}
\affiliation{College of Physical Science and Technology, Sichuan University, Chengdu 610064}
\author{M.~Zeng}
\affiliation{Key Laboratory of Particle and Radiation Imaging (Ministry of Education) and Department of Engineering Physics, Tsinghua University, Beijing 100084}
\author{X.H.~Zeng}
\affiliation{YaLong River Hydropower Development Company, Chengdu 610051}
\author{Z.~Zeng}
\affiliation{Key Laboratory of Particle and Radiation Imaging (Ministry of Education) and Department of Engineering Physics, Tsinghua University, Beijing 100084}
\author{L.~Zhang}
\affiliation{NUCTECH Company, Beijing 10084}
\author{Y.H.~Zhang}
\affiliation{YaLong River Hydropower Development Company, Chengdu 610051}
\author{M.G.~Zhao}
\affiliation{School of Physics, Nankai University, Tianjin 300071}
\author{Z.Y.~Zhou}
\affiliation{Department of Nuclear Physics, China Institute of Atomic Energy, Beijing 102413}
\author{J.J.~Zhu}
\affiliation{College of Physical Science and Technology, Sichuan University, Chengdu 610064}
\author{W.B.~Zhu}
\affiliation{NUCTECH Company, Beijing 10084}
\author{X.Z.~Zhu}
\affiliation{Key Laboratory of Particle and Radiation Imaging (Ministry of Education) and Department of Engineering Physics, Tsinghua University, Beijing 100084}
\author{Z.H.~Zhu}
\affiliation{YaLong River Hydropower Development Company, Chengdu 610051}

\collaboration{CDEX Collaboration}
\noaffiliation


\date{\today}

\begin{abstract}
The CDEX-1 experiment conducted a search of low-mass ($<$ 10 GeV/c$^{2}$) Weakly Interacting Massive Particles (WIMPs) dark matter at the China Jinping Underground Laboratory using a p-type point-contact germanium detector with a fiducial mass of 915 g at a physics analysis threshold of 475 eVee. We report the hardware set-up, detector characterization, data acquisition and analysis procedures of this experiment. No excess of unidentified events are observed after subtraction of known background. Using 335.6 kg-days of data, exclusion constraints on the WIMP-nucleon spin-independent and spin-dependent couplings are derived.
\end{abstract}

\pacs{
95.35.+d,
29.40.-n,
98.70.Vc
}
\keywords{
Dark matter,
Solid-State Detectors,
Background radiation
}

\maketitle
\section{I. Introduction}
\label{1.introduction}
The long-term goal of the CDEX (China Dark matter EXperiment) program~\cite{cdextarget} is to conduct an experiment at the China Jinping Underground Laboratory (CJPL)~\cite{cjpl} with a ton-scale point-contact germanium detector array for low-mass WIMP searches~\cite{rpp,dm2015,dm_dbd} and studies of double-beta decay in $^{76}$Ge~\cite{rpp,dm_dbd,dbd2015,dbd2012}.

The pilot experiment CDEX-0 was with small planar germanium detectors in array form with a target mass of 20 g~\cite{cdex0}, achieving a threshold of 177 eVee (electron equivalent energy eVee is used to characterize detector response throughout in this article, unless otherwise stated). The CDEX-1 experiment adopted kg-scale p-type point contact germanium ($\textsl{p}$PCGe) detectors. Data taking of the first phase was performed only with a passive shielding system, and dark matter results were published with 14.6 kg-days of data taken from August to September, 2012 and a threshold of 400 eVee~\cite{cdex1}. Starting November 2013, Phase-II measurements are based on the design of earlier work~\cite{texono2003-2007,cdex0}, with an active NaI(Tl) anti-Compton (NaI-AC) detector installed. First results with 53.9 kg-days of data were reported~\cite{cdex12014}, providing an order of magnitude improvement on the spin-independent $\chi$-N coupling (WIMPs denoted by $\chi$). In particular, the allowed region implied by the CoGeNT~\cite{cogent} experiment is probed and excluded with an identical detector target.

We describe the details of the CDEX-1 experiment and report the results with 335.6 kg-days of data taking at CJPL in the following Sections.

\section{II. Experimental Setup}
\label{2.experimental setup}
\subsection{A. China Jinping Underground Laboratory}
\label{2.1 CJPL introduction}
The China Jinping Underground Laboratory (CJPL) is located in Sichuan province, with a vertical rock overburden of more than 2400 m, providing 6720 meters of water equivalent overburden as passive shield against cosmic rays and their induced backgrounds. The flux of cosmic ray and associated backgrounds is down to 61.7 y$^{-1}$$\cdot$m$^{-2}$~\cite{cjplmuon}. In addition, the radioactivities of $^{232}$Th, $^{238}$U and $^{40}$K from rock surrounded CJPL were very low based on in situ measurement~\cite{cjplgamma}. The low cosmic-ray flux and radioactivities of $^{238}$U and $^{232}$Th give rise to low level of neutron flux.
\subsection{B. Detector Hardware}
\label{2.2 DAQ}
CDEX-1 experiment adopted one single module at 1kg-scale mass $\textsl{p}$PCGe to search for WIMPs. The p-type germanium crystal is a cylinder with about 62 mm of both height and diameter which give rise to 994 g mass. It has two electrodes, the outer electrode is n$^{+}$ type, providing high voltage (HV) and signal, and the tiny point-like center electrode is p$^{+}$ type, with order of 1 mm diameter resulting in order of 1 pF capacitance, leading to low energy threshold in potential. At phase I experiment, the outer electrode signal was read out by a resistive feedback preamplifier~\cite{cdex1}. At phase II measurement, the signal output was removed due to its induced noise, such that the outer electrode served only as a HV electrode. The center electrode signal was read out by an ultra-low noise JFET nearby and then supplied into a pulsed-reset feedback preamplifier. The preamplifier generates three identical energy-related signals (OUT$_{-}$E), one timing-related signal (OUT$_{-}$T) and one inhibit signal (IHB) marking the inactive time of the preamplifier. Meanwhile, the preamplifier can accept a test input, typically from an electronic pulser to simulate physical signals.

\begin{figure}[t] 
\centering
\includegraphics[height=5.5cm,width=8.0cm]{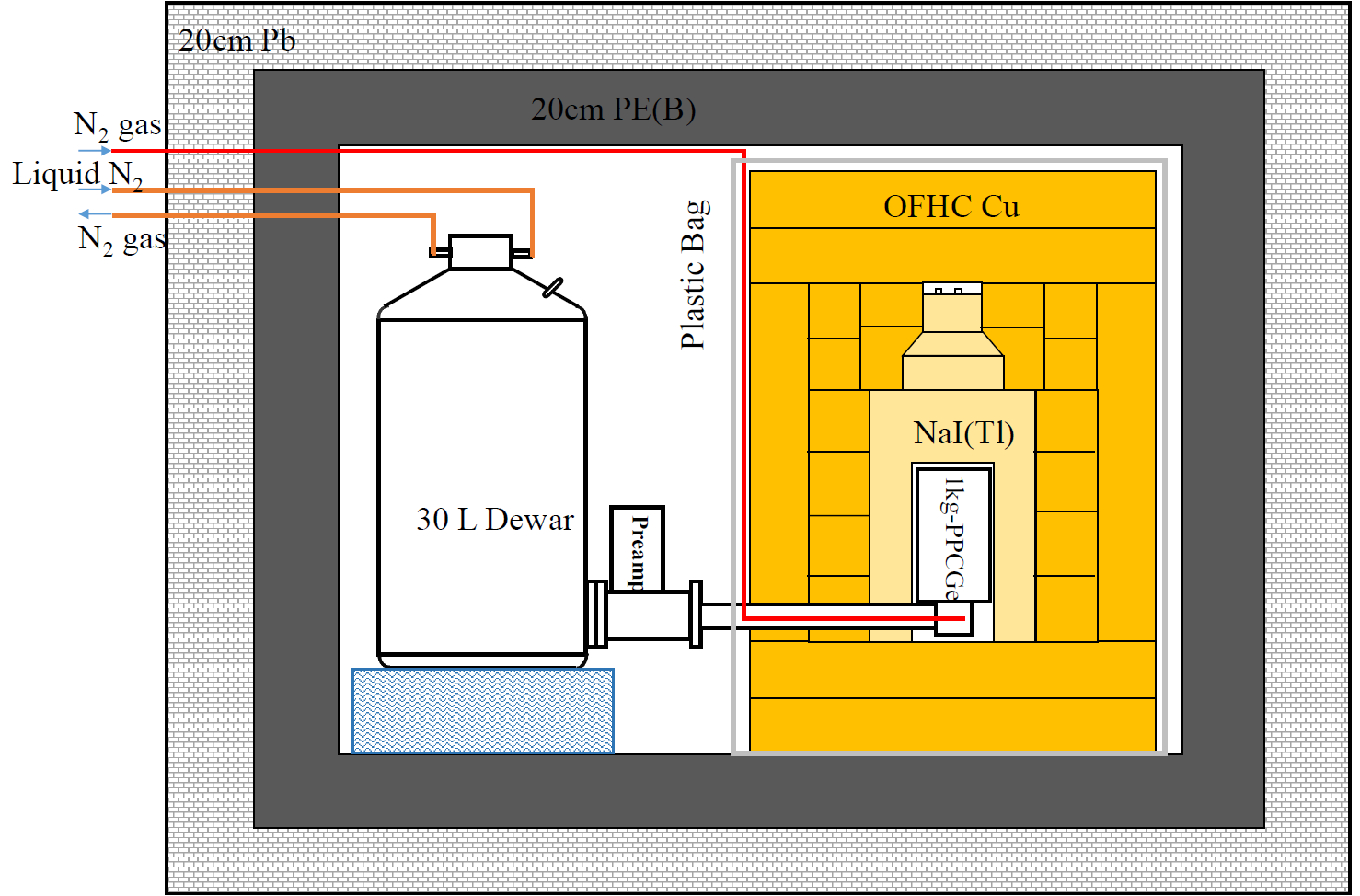}
\caption{\label{fig1} Schematic diagram of CDEX-1 experimental setup}
\end{figure}

The NaI(Tl) scintillator crystal of the AC detector is well-shaped which can enclose the cryostat of the $\textsl{p}$PCGe, as shown in Figure~\ref{fig1}, and the thickness of its side and top is 48 mm and 130 mm, respectively. The scintillation light from NaI(Tl) crystal were read out by a photomultiplier tube (PMT), which has two outputs from anode and dynode respectively, one was loaded to a shaping amplifier at high gain determining the time over NaI-AC energy threshold, and another was loaded to a timing amplifier at low gain, which was used to measure energy as well as discriminate background sources based on pulse characteristics for different radiation.

\begin{figure*}[htbp] 
\includegraphics[height=9.0cm,width=17.0cm]{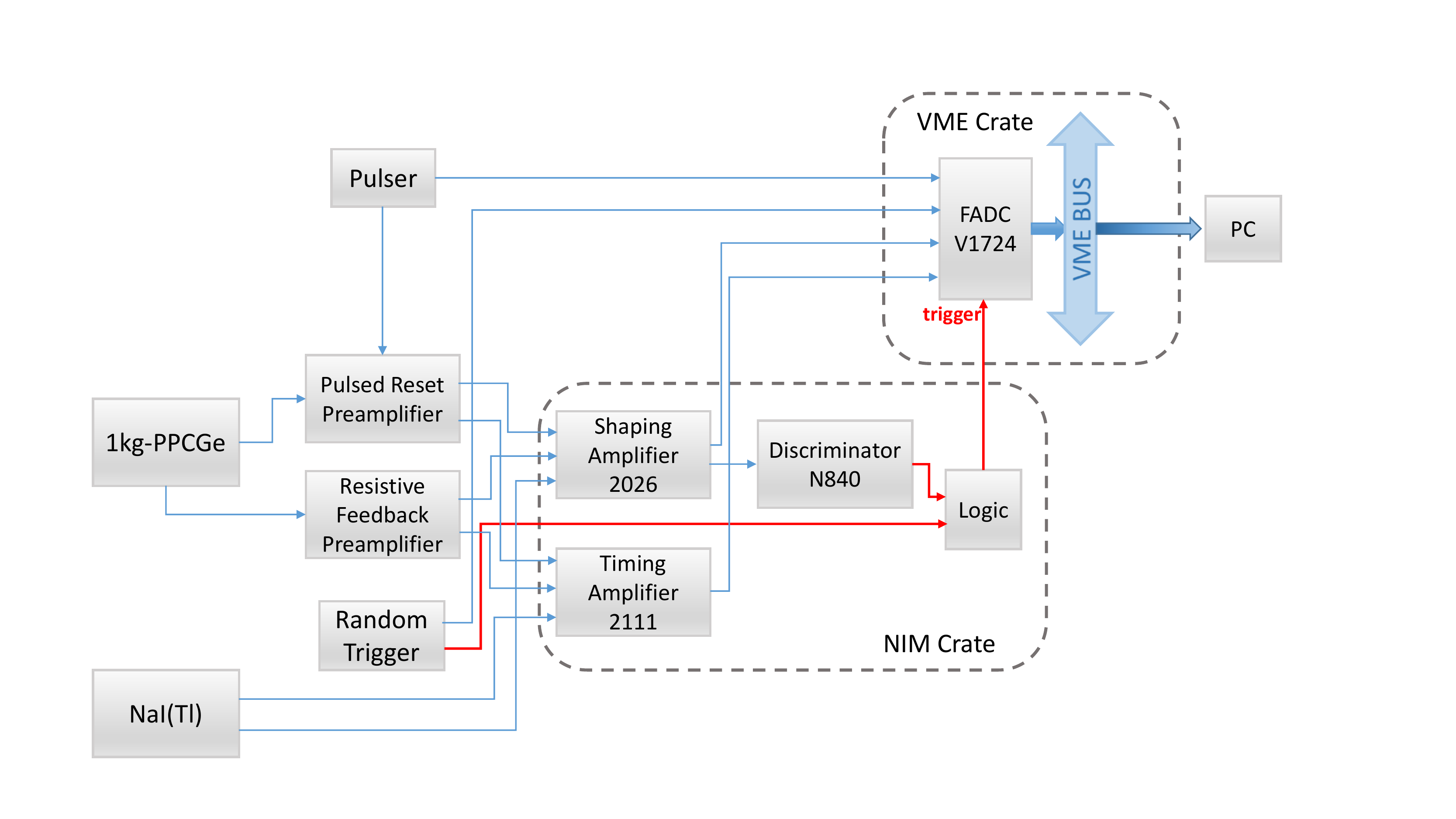}
\caption{\label{fig2} Schematic diagram of the data acquisition system for CDEX-1 phase II.}
\end{figure*}

The schematic of CDEX-1 data acquisition (DAQ) system is shown in Figure~\ref{fig2}, which was based on commercial NIM/VME modules and crates from CANBERRA and CAEN. The $\textsl{p}$PCGe worked at +3500 V provided by an high voltage module (CANBERRA 3106D). The three identical OUT$_{-}$E signals were loaded to shaping amplifiers (CANBERRA 2026) at 6 us (S$_{\textrm{p}6}$), 12 us shaping time (S$_{\textrm{p}12}$) and a timing amplifier (CANBERRA 2111) (T$_{\textrm{p}}$), respectively. Each gain of these amplifiers was adjusted to achieve maximal signal-to-noise ratio and maximal information for low energy events. The energy range was limited to 12 keVee. The S$_{\textrm{p}6,12}$ signal provided energy measurement and system trigger of the data acquisition (DAQ). The T$_{\textrm{p}}$ signal recorded the raw pulse shape information of one event, so it can provide the rise time information. The OUT$_{-}$T signal was distributed into a timing amplifier with low gain to measure high energy backgrounds, intending to analysis background source and opening a window to study $^{76}$Ge neutrinoless double-beta decay. These outputs were digitized and recorded by a flash analog-to-digital convertor (FADC, CAEN V1724) at 100 MHz sampling rate with a resolution of 14 bit. The data acquisition software is based on LabVIEW program. The discriminator output of the inhibit signal provided another trigger of the DAQ and was recorded to determine the exact time of the beginning of discharge process of the preamplifier. To monitor the noise level and dead time of the system, random trigger signals (RT) at 0.05 Hz generated by a precision pulser were injected into the system, providing system trigger.

\begin{figure}[ht] 
\centering
\includegraphics[width=1.0\linewidth]{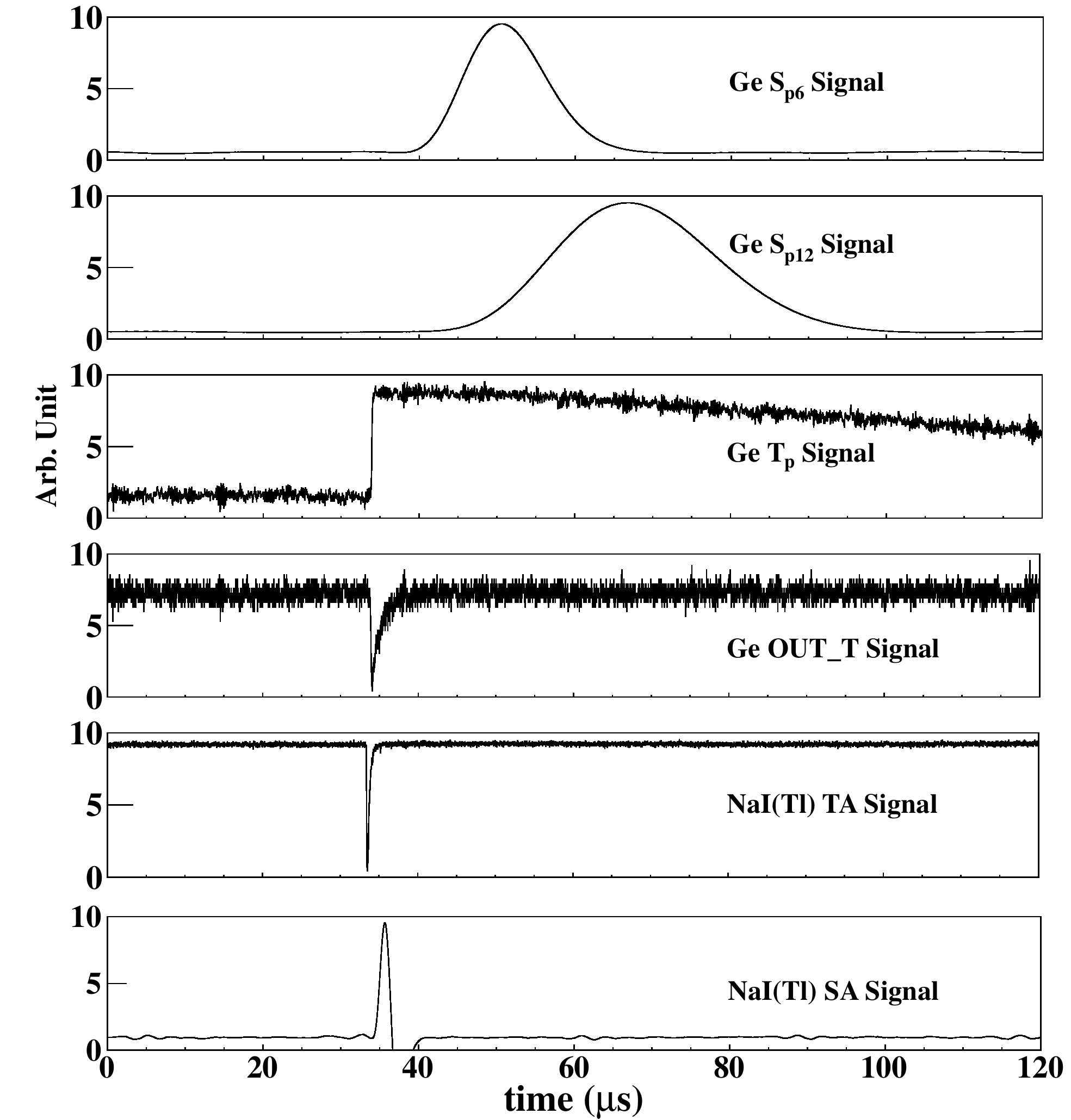}
\caption{\label{fig3} Example of one AC$^{+}$ event recorded by FADC, corresponding to energy $\sim$10.37 keVee deposited in $\textsl{p}$PCGe.}
\end{figure}

The NaI-AC detector is optimized on its energy threshold, energy linearity in broad energy range, energy resolution and stability. The NaI-AC signals were recorded only when the $\textsl{p}$PCGe detector was fired and triggered the DAQ, and this kind of coincidence events was denoted as AC$^{+}$. The anticoincidence events which only fired in the $\textsl{p}$PCGe detector but without signals at the NaI-AC detector are denoted as AC$^{-}$. Figure~\ref{fig3} shows an example of AC$^{+}$ event recorded by the DAQ. In general, the DAQ took data at low trigger rate ($\sim$3-5 Hz) to decrease penalty of dead time.

\subsection{C. Shielding System}
\label{2.3 shield}
The passive shielding structure of CDEX-1 in CJPL is displayed in Figure~\ref{fig1}. The outermost is 20 cm of lead to shield ambient gamma ray. The inner is 20 cm thick layer of $\sim$ 30$\%$ borated polyethylene, acting as thermal neutron absorber. At phase I experiment, the innermost is a minimum of 20 cm of Oxygen Free High Conductivity (OFHC) copper surrounding the 994 g $\textsl{p}$PCGe detector cryostat in all directions, to further reduce gamma ray surviving from outer shield. Exterior to the OFHC shield is a plastic bag which is used to seal the working space to prevent radon incursion. The radon exclusion volume is continuously flushed with nitrogen gas from a pressurized Dewar. At phase II experiment, interior to OFHC shield is a NaI-AC detector with a well-shaped cavity enclosing the $\textsl{p}$PCGe detector cryostat to provide passive and active shielding. Detailed discussion about its performances is provided in Section III. The entire structure was located in a 1 m thick of polyethylene room, which can moderate and absorb ambient neutron.

\begin{figure}[ht] 
\centering
\includegraphics[width=1.0\linewidth]{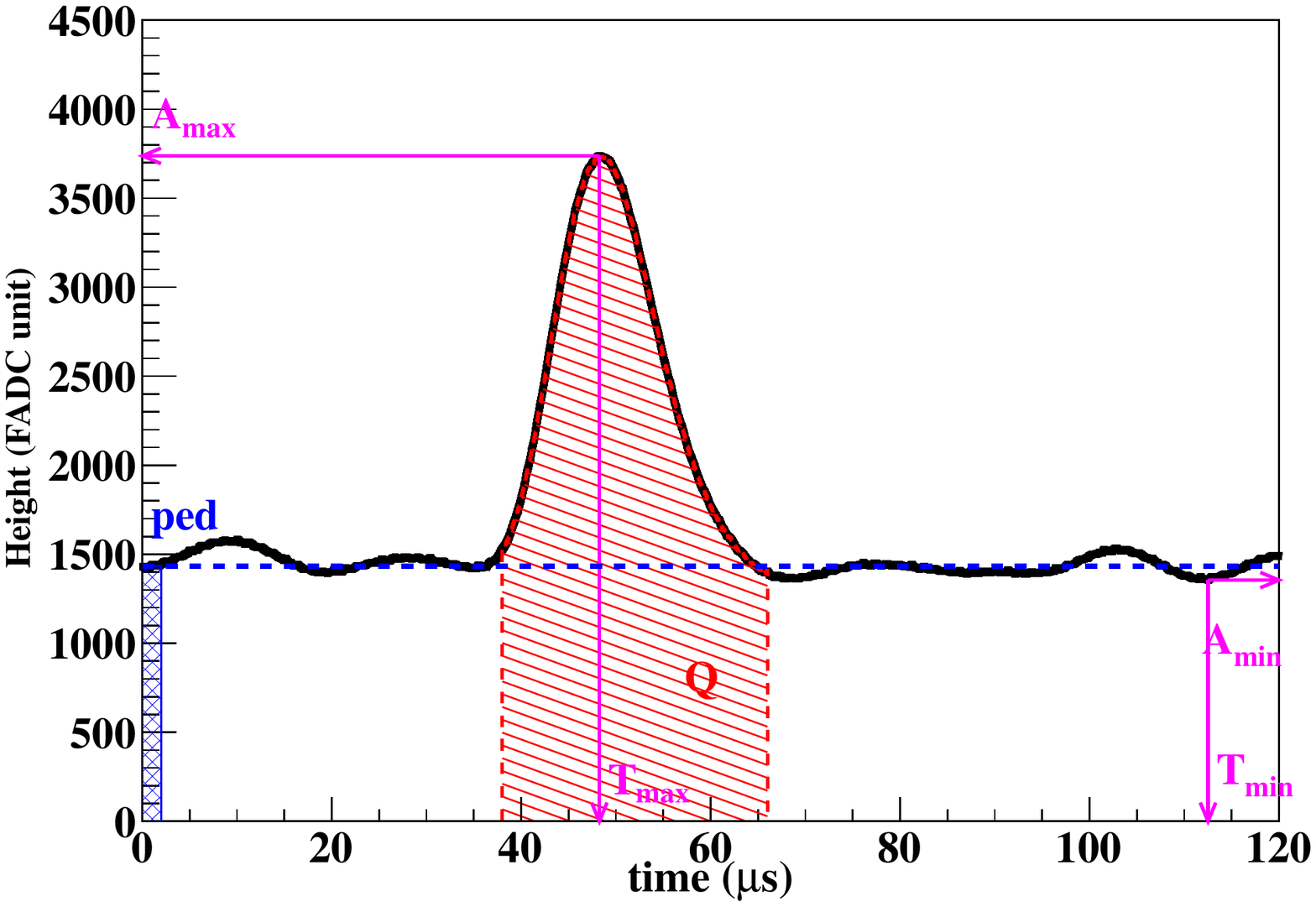}
\caption{\label{fig4} Typical pulse of S$_{\textrm{p}6}$ at 1.84 keVee. Some parameters are defined, (A$_{\textrm{max}}$ , T$_{\textrm{max}}$) represent the maximal amplitude and its corresponding time of the pulse; (A$_{\textrm{min}}$ , T$_{\textrm{min}}$) represent the minimal amplitude and its corresponding time of the pulse; Q means integration of the pulse; Ped means the pedestal of the pulse.}
\end{figure}

\section{III. Detector Characterization}
\label{3.ppcge characterization}
The performances of the detection system were studied in details. Characterization of the $\textsl{p}$PCGe, the NaI-AC and the DAQ are discussed in the following Sections.
\subsection{A. Energy Definition and Calibration}
\label{3.1 energy calibration}
\begin{figure}[ht] 
\centering
\includegraphics[width=8.0cm,height=7.0cm]{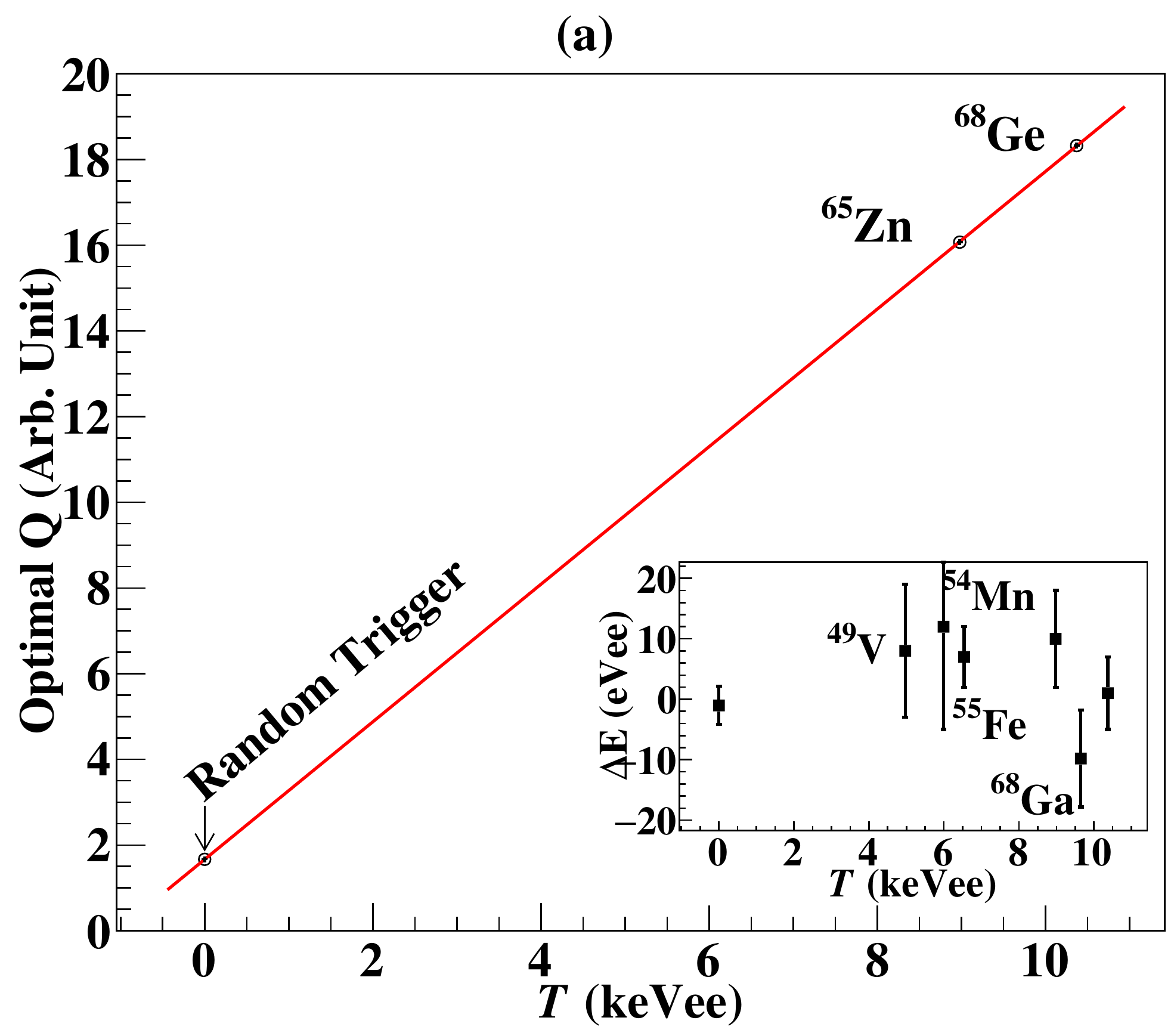}
\includegraphics[width=8.0cm,height=7.0cm]{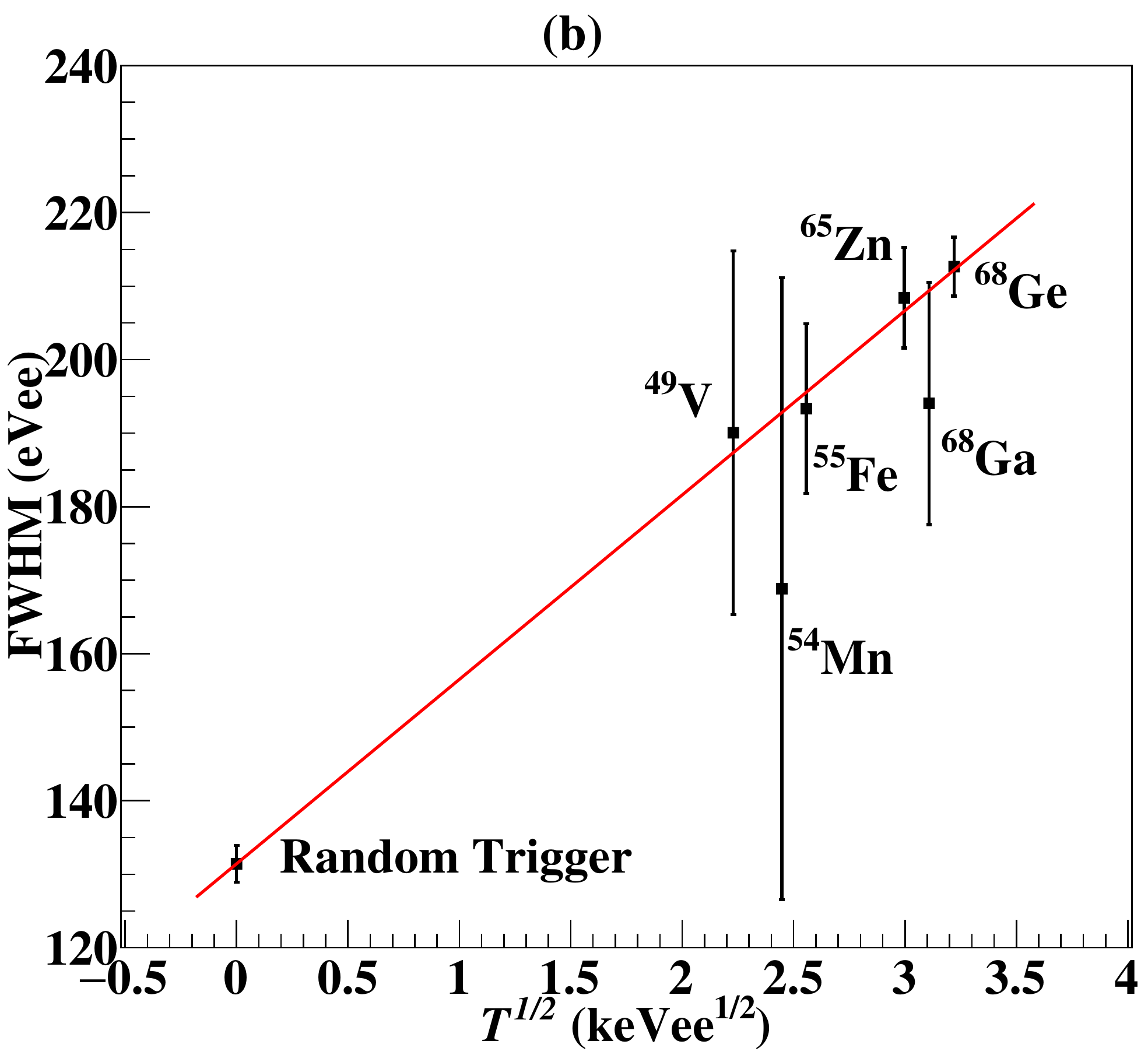}
\caption{\label{fig5} (a) Calibration line relating the optimal Q measurements from S$_{\textrm{p}6}$ with the known energies from $^{68}$Ge and $^{65}$Zn K-shell X rays and RT. The error bars are smaller than the data point size. The energy difference between the energy derived from the calibration and the real energy for these three peaks are depicted in the inset, together with K-shell X rays of $^{68}$Ga, $^{55}$Fe, $^{54}$Mn and $^{49}$V. (b) Relation between energy of K-shell X rays and energy resolution.}
\end{figure}

A typical pulse of the $\textsl{p}$PCGe was displayed in Figure~\ref{fig4}, with the parameters defined consistently for all channels. Two energy-related parameters are defined: (i) maximal amplitude of one pulse (A$_{\textrm{max}}$); (ii) integration of one pulse (Q). Optimized partial integration of S$_{\textrm{p}6}$ was chosen to define as energy ($\textsl{T}$) for its excellent energy linearity at low energy range. Since the active volume of $\textsl{p}$PCGe crystal is surrounded by $\sim$ 1.0 mm dead layer and 1.5 mm of OFHC copper cryostat, external low energy X-rays at at $<$ 50 keVee range cannot penetrate into the $\textsl{p}$PCGe crystal. Energy calibration was therefore done with its internal characteristic X-rays originated from the electron capture (EC) of the cosmogenic radioisotopes~\cite{texono2013,cdex1,cdex12014}. Figure~\ref{fig5}(a) shows the energy calibration by the two dominant K-shell X-rays: $^{68}$Ge (10.368 keVee), $^{65}$Zn (8.98 keVee) and RT events (0 keVee). The inset figure displays energy difference between the calibrated energy and the real energy of these three peaks, together with other peaks observed in the measured CDEX-1 background spectrum, demonstrating good linearity of less than 0.8$\%$ deviation. The relationship between energy and its resolution is also depicted in Figure~\ref{fig5}(b), showing good linearity between $\surd$\textit{T} and the energy resolution FWHM (Full Width at Half Maximum). The energy resolution at low energy region is derived from this line.

\begin{figure}[ht] 
\centering
\includegraphics[height=7.0cm,width=8.0cm]{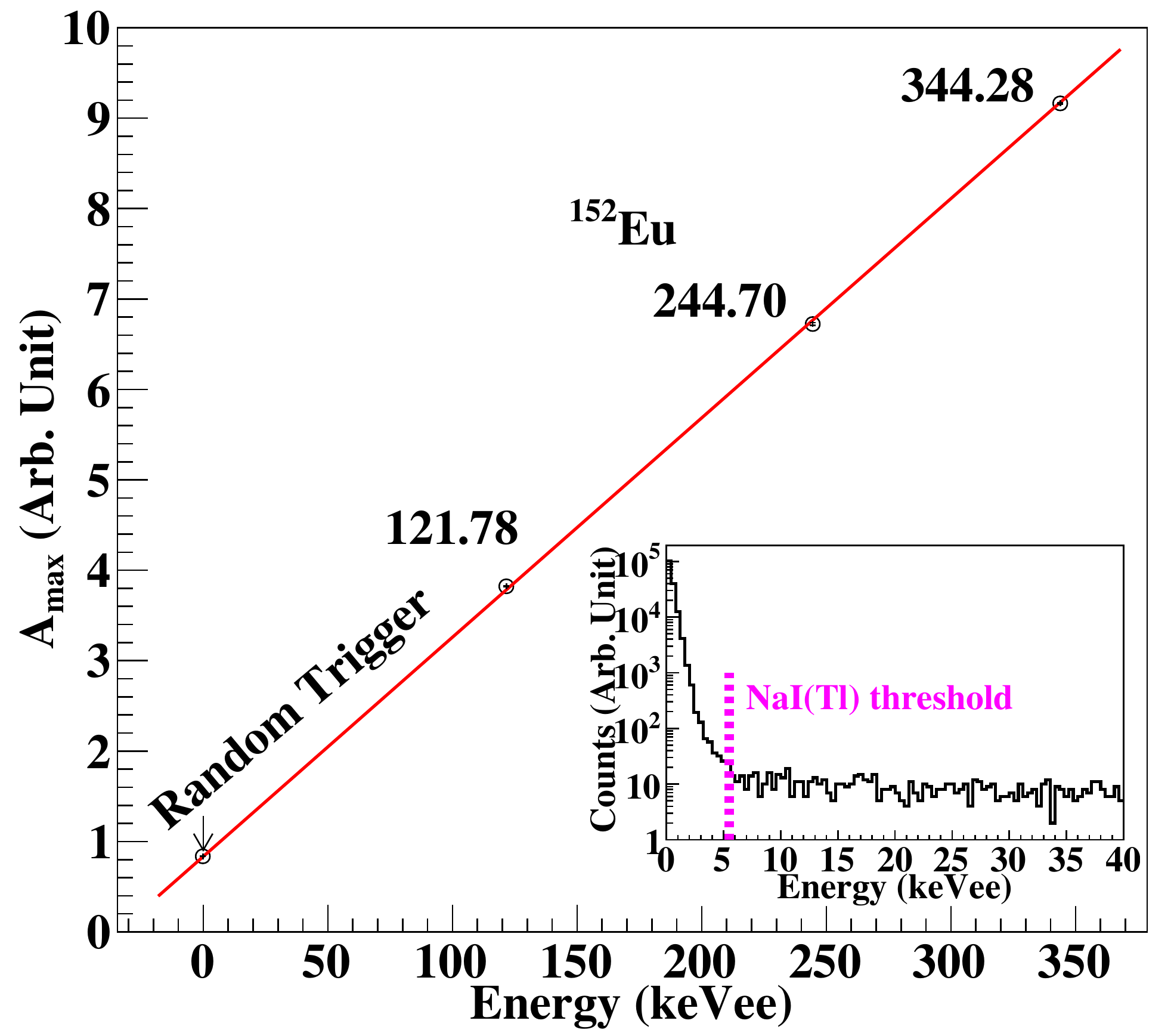}
\caption{\label{fig6} Energy calibration for NaI(Tl) SA channel. The error bars are smaller than the data point size. The inset figure depicts the measured background energy spectrum of NaI(Tl), and the energy threshold was set at the edge of noise.}
\end{figure}

NaI-AC detector was developed with emphasis on low energy threshold to achieve high efficiency of AC$^{+}$ background suppression. A$_{\textrm{max}}$ was used to define its energy, and calibrated by a $^{152}$Eu (121.78 keV, 244.70 keV, 344.28 keV) source together with RT events. The energy threshold of NaI-AC detector was achieved as low as 6 keVee for background measurement, as illustrated in Figure~\ref{fig6}.

\subsection{B. Quenching Factor}
\label{3.2 QF}
\begin{figure}[t] 
\centering
\includegraphics[height=8.0cm,width=8.0cm]{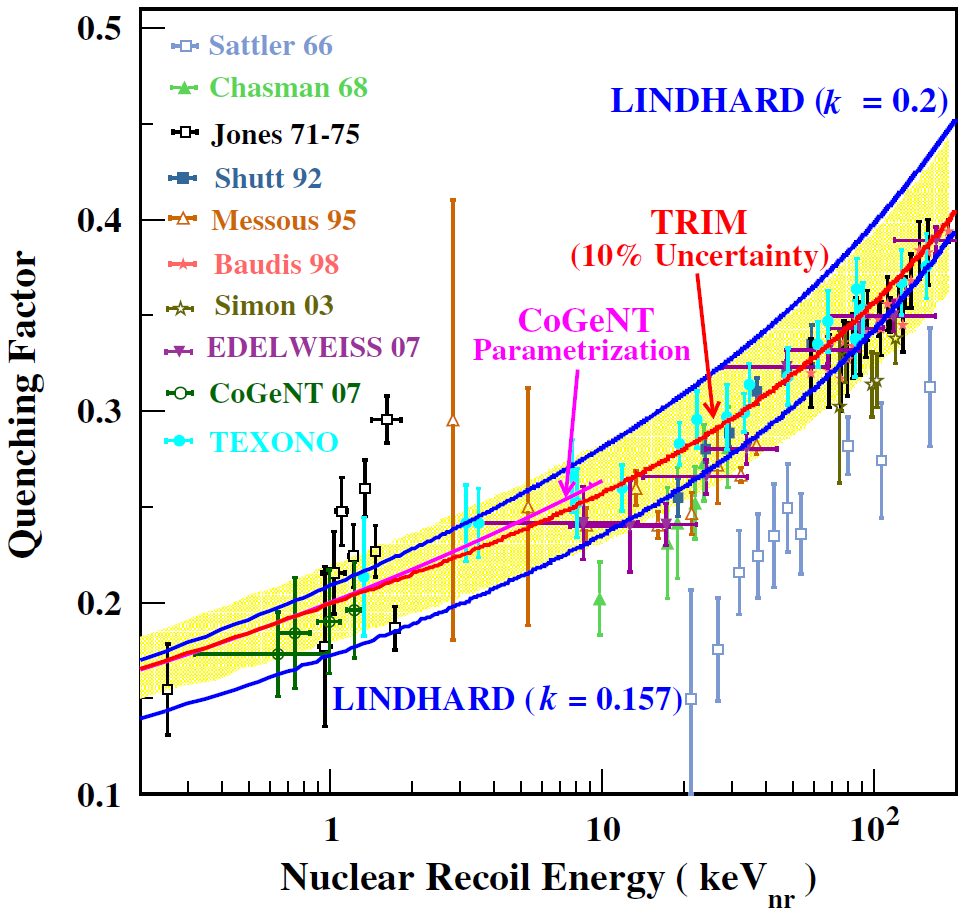}
\caption{\label{fig7} QF results for germanium from both experiments and calculations. The QF curve which is derived from TRIM~\cite{trim} as a function of nuclear recoil energy, together with a 10$\%$ systematic error band. The various experimental measurements are overlaid, so are the alternative QFs from parameterization of CoGeNT~\cite{cogent} and the Lindhard theory~\cite{lindard} at \textit{k}=0.2 and \textit{k}=0.157 adopted by CDMSlite~\cite{cdmslite}. It shows that the TRIM results with uncertainties covers most data points as well as the alternative formulations.}
\end{figure}

Quenching factor (QF) is defined as the ratio of the measured energy to the total nuclear recoil energy deposited in the detector medium. It is crucial to know the relation between QF and nuclear recoil energy in the studies of WIMP search. Figure 7 showed a compilation of all experiment measurements and calculations of QF for recoiled germanium nuclei~\cite{cdex12014}. Several experiments have measured the QF down to a few keVnr (nuclear recoil energy). Typically, two methods can be used to calculate the QF for different nuclear recoil energy. In TRIM software simulation, several aspects of stopping power, range and straggling distributions of a recoiled nucleon with certain energy are considered, while Hartree-Fock atoms and lattice effects are also included~\cite{trim}. In analytic Lindhard calculation, an ideal and static atom is adopted, and Lindhard model is parameterized to a constant \textit{k} which is related to stopping power~\cite{lindard}. The TRIM results agree well with the QF experimental results at a larger energy range and therefore are adopted in our analysis. As illustrated in Figure~\ref{fig7}, QF function derived from TRIM with a 10$\%$ systematic uncertainty is applied in our analysis.

\subsection{C. Dead Layer}
\label{3.3 dead layer}
\begin{figure}[ht] 
\centering
\includegraphics[height=6.0cm,width=8.0cm]{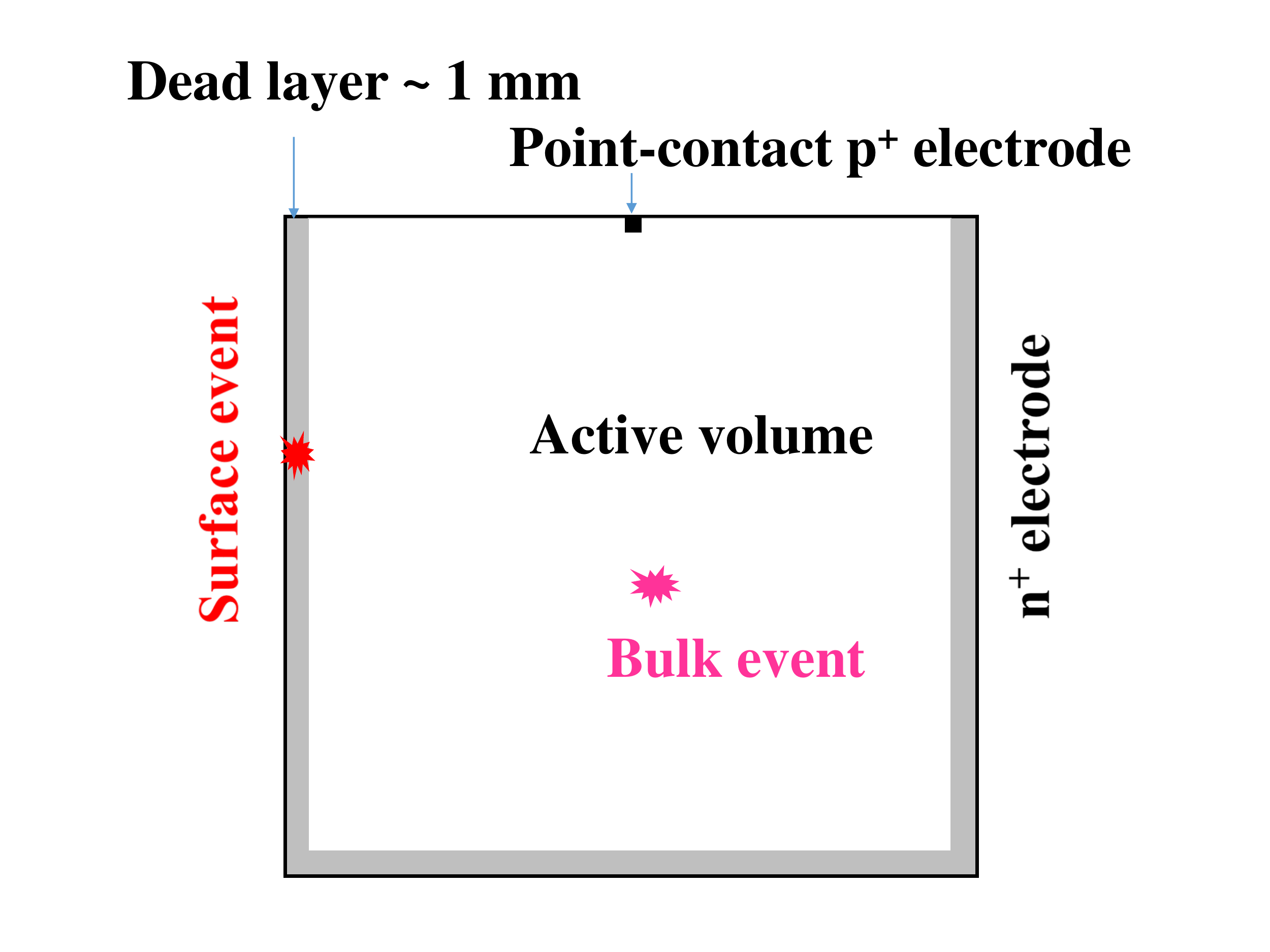}
\caption{\label{fig8} Schematic diagram of $\textsl{p}$PCGe crystal configuration.}
\end{figure}

The n$^{+}$ outer surface electrode of $\textsl{p}$PCGe is fabricated by lithium diffusion, resulting in normally about 1 mm depth of dead layer. This dead layer is composed of totally dead layer where the electric field is zero, and transition layer where the electric field is weak. Interior to transition layer is active volume. Electron-hole pairs generated from events taking place in transition layer have slower drift velocity than those in active volume, leading to pulse with typically slow rise time as well as degraded amplitude due to partial charge collection~\cite{texonobs}. We denoted events at the active volume with complete charge collection as bulk events and events at the dead layer as surface events, as illustrated in Figure~\ref{fig8}. The totally dead layer acts as passive shield against external low energy $\gamma$/$\beta$, and transition layer acts as active shield against ambient gamma rays through bulk/surface events discrimination based on rise time characteristics. This is self-shield effect of $\textsl{p}$PCGe. On the contrary, the dead layer produced fiducial mass loss. Since the attenuation of gamma rays by the dead layer was dependent on energy, the ratio of these gamma rays at photoelectron peaks would be changed. $^{133}$Ba source with various energy gamma rays was used to measure the thickness of the dead layer for the $\textsl{p}$PCGe, and it was derived to be (1.02$\pm$0.14) mm via comparison of measured and simulated intensity ratios of those gamma peaks~\cite{deadlayer_majorana}. This give rise to fiducial mass to be 915 g with 1$\%$ uncertainty.

\subsection{D. Trigger Efficiency}
\label{3.4 trigger}
\begin{figure}[t] 
\centering
\includegraphics[height=7.0cm,width=8.0cm]{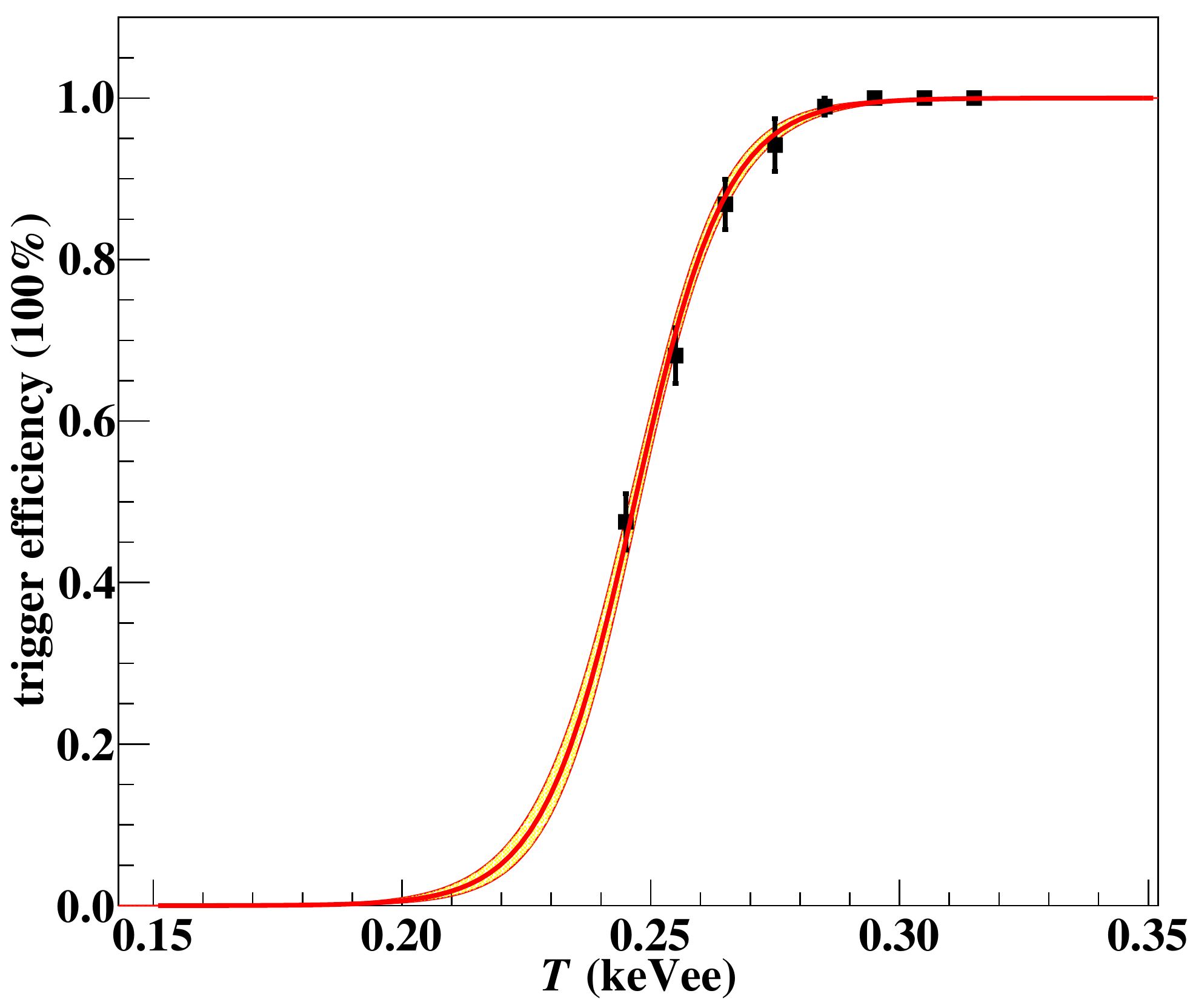}
\caption{\label{fig9} Trigger efficiency derived from $^{137}$Cs AC$^{+}$ samples, adopting error function to fit the experiment data points, and 1 $\sigma$ band of the error was superimposed.}
\end{figure}

In principle, physical events over the DAQ threshold would produce triggers and be recorded. The efficiency that events produced triggers for the DAQ is defined as trigger efficiency, which was 50$\%$ for events at the discriminator threshold. AC$^{+}$ events from source sample were used to derive the trigger efficiency~\cite{cdex0,texono2009,cdex1}. Figure~\ref{fig9} displayed the trigger efficiency together with 1$\sigma$ band derived from $^{137}$Cs AC$^{+}$ sample. It is shown that the trigger threshold was 246$\pm$2 eVee, and the trigger efficiency was 100$\%$ above our analysis threshold 475 eVee.

\subsection{E. Stability}
\label{3.5 stability}
\begin{figure}[t] 
\centering
\includegraphics[width=1.0\linewidth]{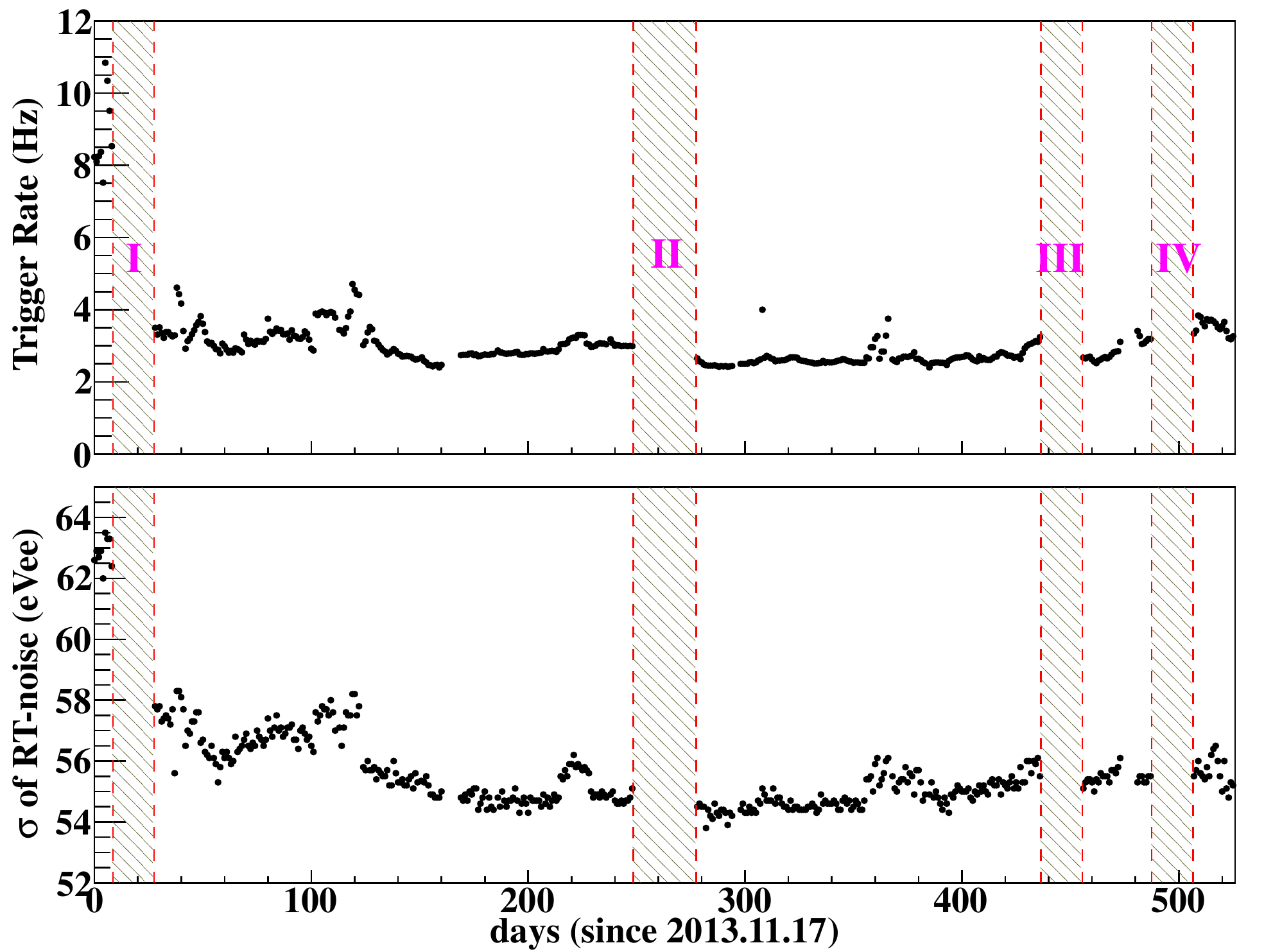}
\caption{\label{fig10} Top Panel: daily average trigger rate of the $\textsl{p}$PCGe detector system; Bottom Panel: daily average RT electronic noise of the $\textsl{p}$PCGe detector system.}
\end{figure}

Both the trigger rate and the noise of RT of the $\textsl{p}$PCGe detector were monitored, shown in Figure~\ref{fig10}. An improvement of the Laboratory power supply took place at the time period of I. A power filter was used to stabilize the power supply, and the electronic noise of the detector system decreased around 10$\%$. Calibration was performed from late July to late August, 2014, corresponding to the time period of II. During the time period of III and IV, the construction work at the PE room prevented data taking. Both the trigger rate and the noise of RT were kept stable to 16$\%$ and 2$\%$, respectively, during the data taking periods.

\section{IV. Data Analysis}
\label{4. data analysis}
The data analysis is based on timing and amplitude parameters extracted from pulses recorded by the DAQ described in Section II-B.
\subsection{A. Parameters Definition}
\label{4.1 parameters}
The amplitude parameters are defined in Section III-A. The timing parameters can be classified into three categories: (i) the timing differences between one event and its closest prior and post IHB events, denoted as T$_{-}$ and T$_{+}$, the detailed information described in~\cite{cdex1}; (ii) the timing interval of one event recorded by the $\textsl{p}$PCGe detector and the NaI-AC detector, $\Delta$t; (iii) the rise time of one event $\tau$, defined as the time interval between 5$\%$ and 95$\%$ of the T$_{\textrm{p}}$ pulse height. To calculate the $\tau$, the pulse-processing algorithm in~\cite{texono2013,cdex12014,texonobs} was applied. This rise-time provides the location information where one event happened, in active volume or in dead layer, to discriminate the Bulk/Surface events.

\begin{figure}[t] 
\centering
\includegraphics[height=7.0cm,width=8.5cm]{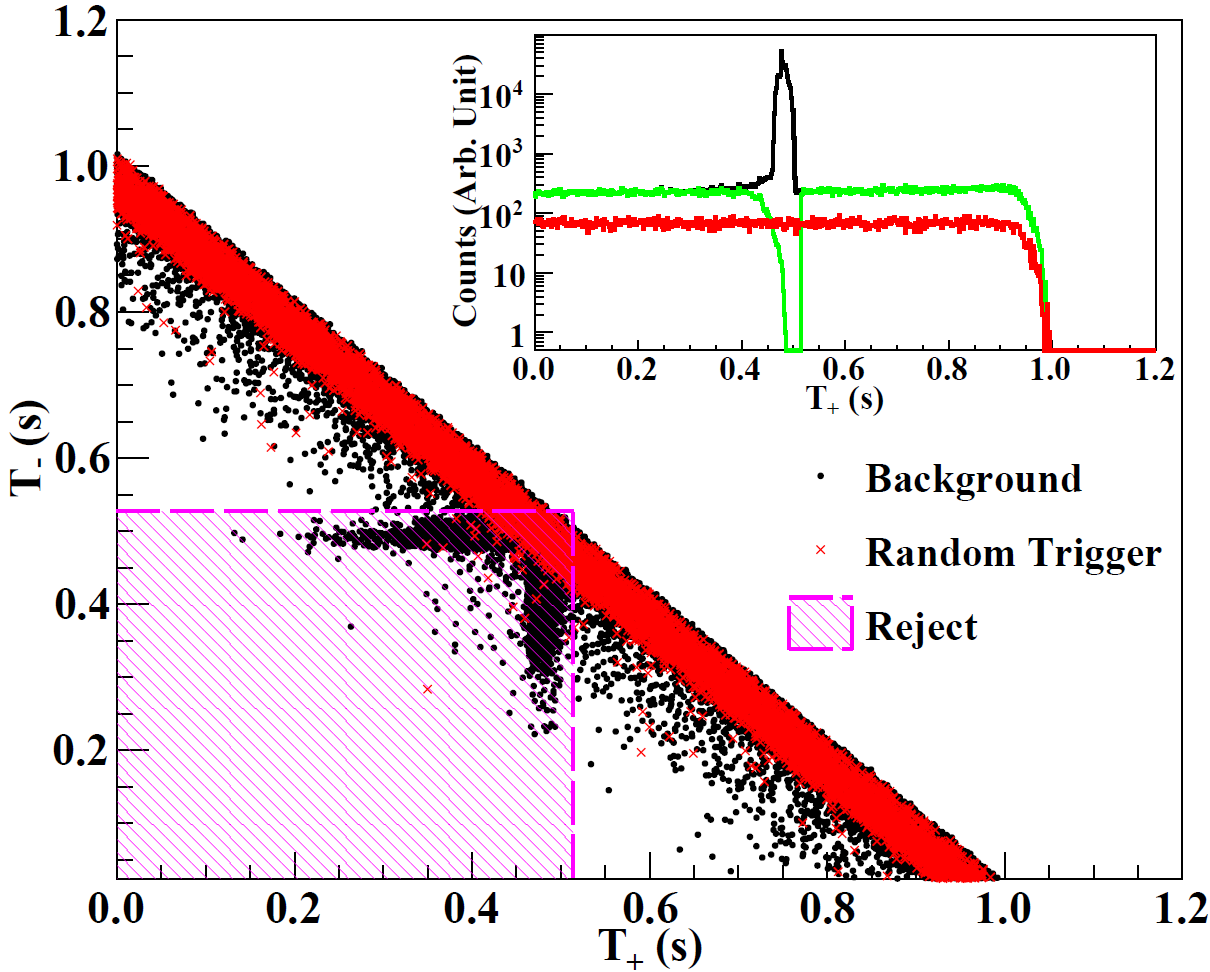}
\caption{\label{figtt} The scatter plots of T$_{-}$ and T$_{+}$ for random trigger events and background events. The TT cut has also been overlaid on the scatter plot. The inset figure shown the T$_{+}$ spectra of background before (black) and after the TT cut (green), together with the T$_{+}$ spectra for RT events (red).}
\end{figure}

\begin{figure}[!ht] 
\centering
\includegraphics[height=6.0cm,width=8.0cm]{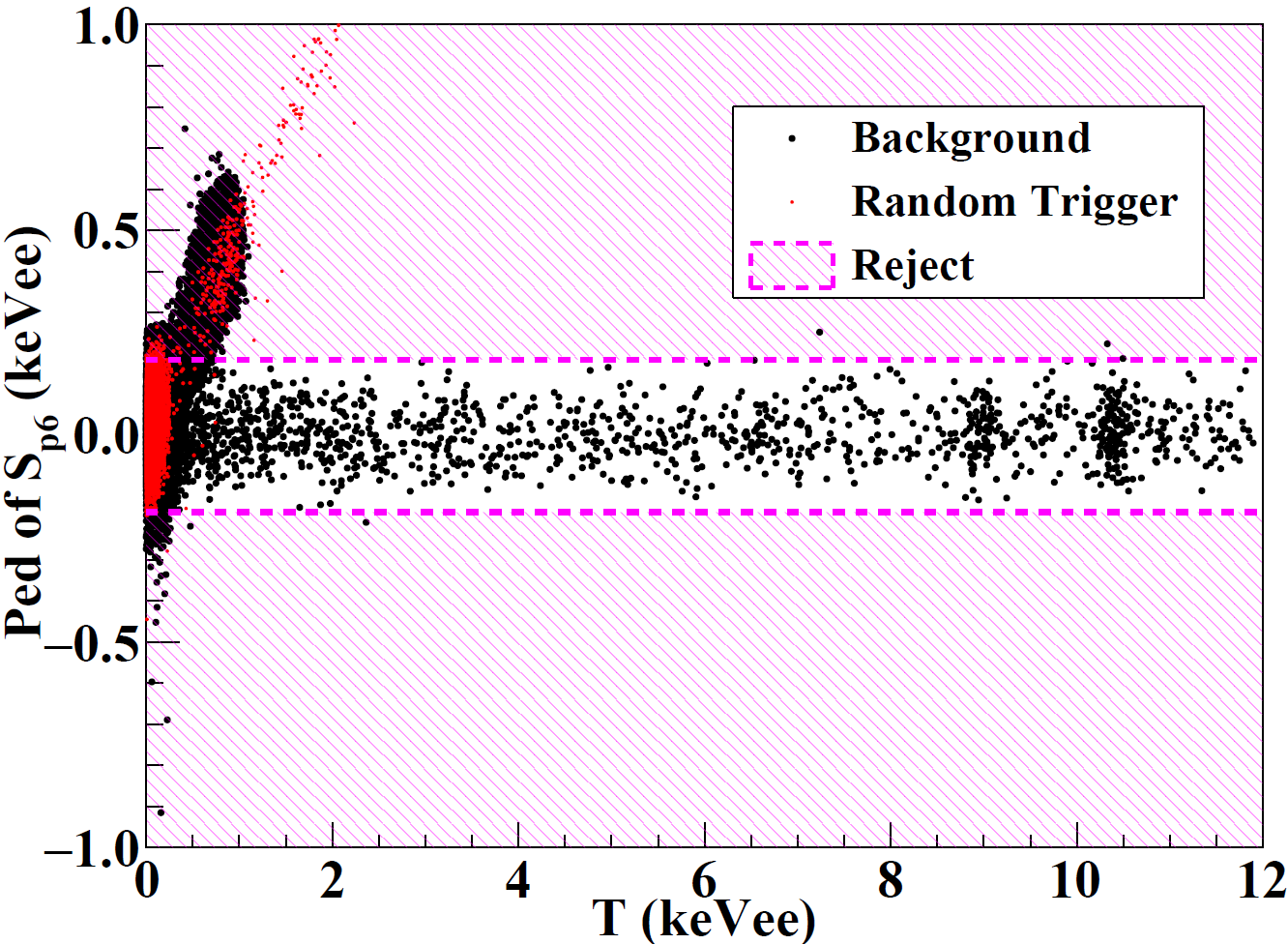}
\caption{\label{fig11ped} The distributions of Ped of S$_{\textrm{p}6}$ for both background and RT events, and the Ped cut criteria.}
\end{figure}

\begin{figure*}[t] 
\includegraphics[height=5.5cm,width=16.5cm]{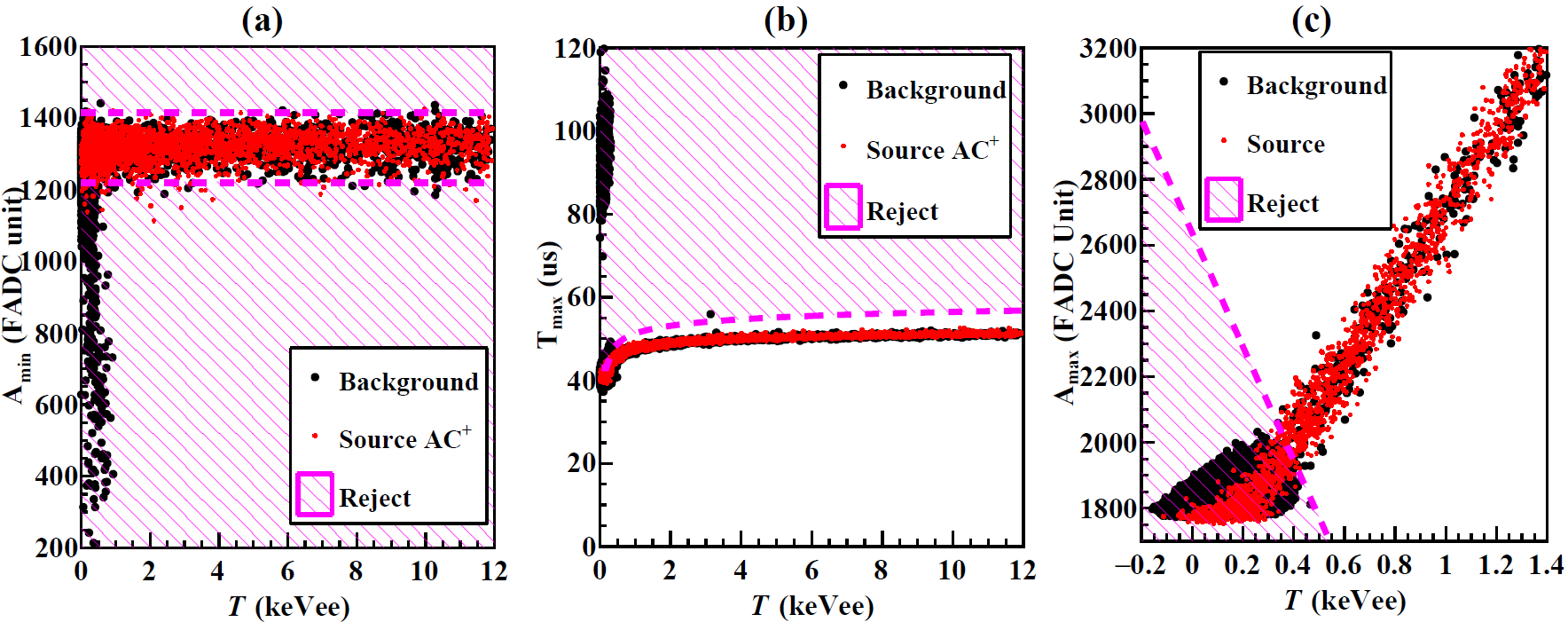}
\caption{\label{fig12bc} The energy dependent PSD cuts: (a) A$_{\textrm{min}}$ cut, (b) A$_{\textrm{max}}$-T$_{\textrm{max}}$ correlation cut, (c) A$_{\textrm{max}}$-Q correlation cut.}
\end{figure*}

\subsection{B. Data Selection}
\label{4.2 data selection procedure}
\begin{figure}[t] 
\centering
\includegraphics[height=6.5cm,width=8.0cm]{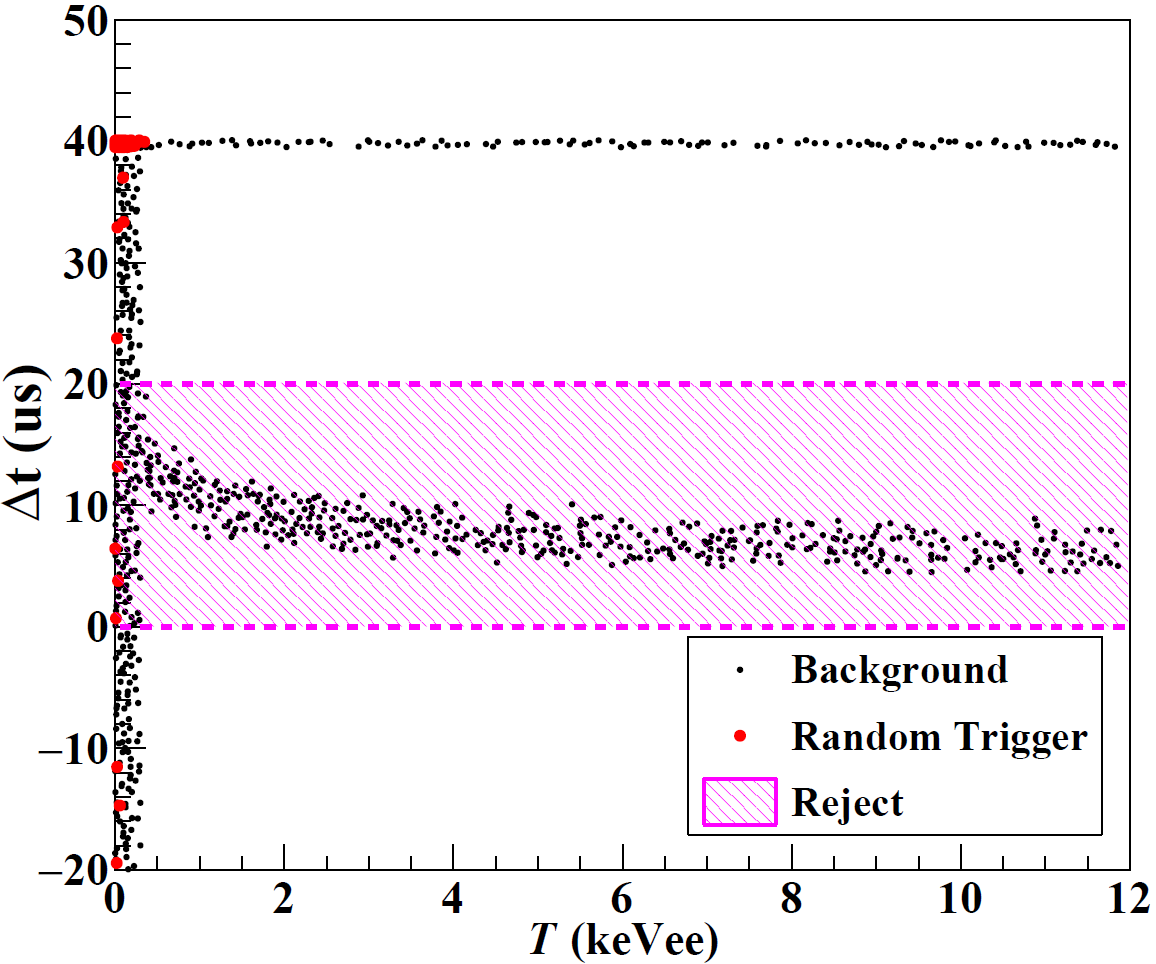}
\caption{\label{fig13ac} $\Delta$t versus energy distribution and AC cut criteria. The rejected band are the AC$^{+}$ events. }
\end{figure}

\begin{figure}[t] 
\centering
\includegraphics[width=8.0cm,height=6.5cm]{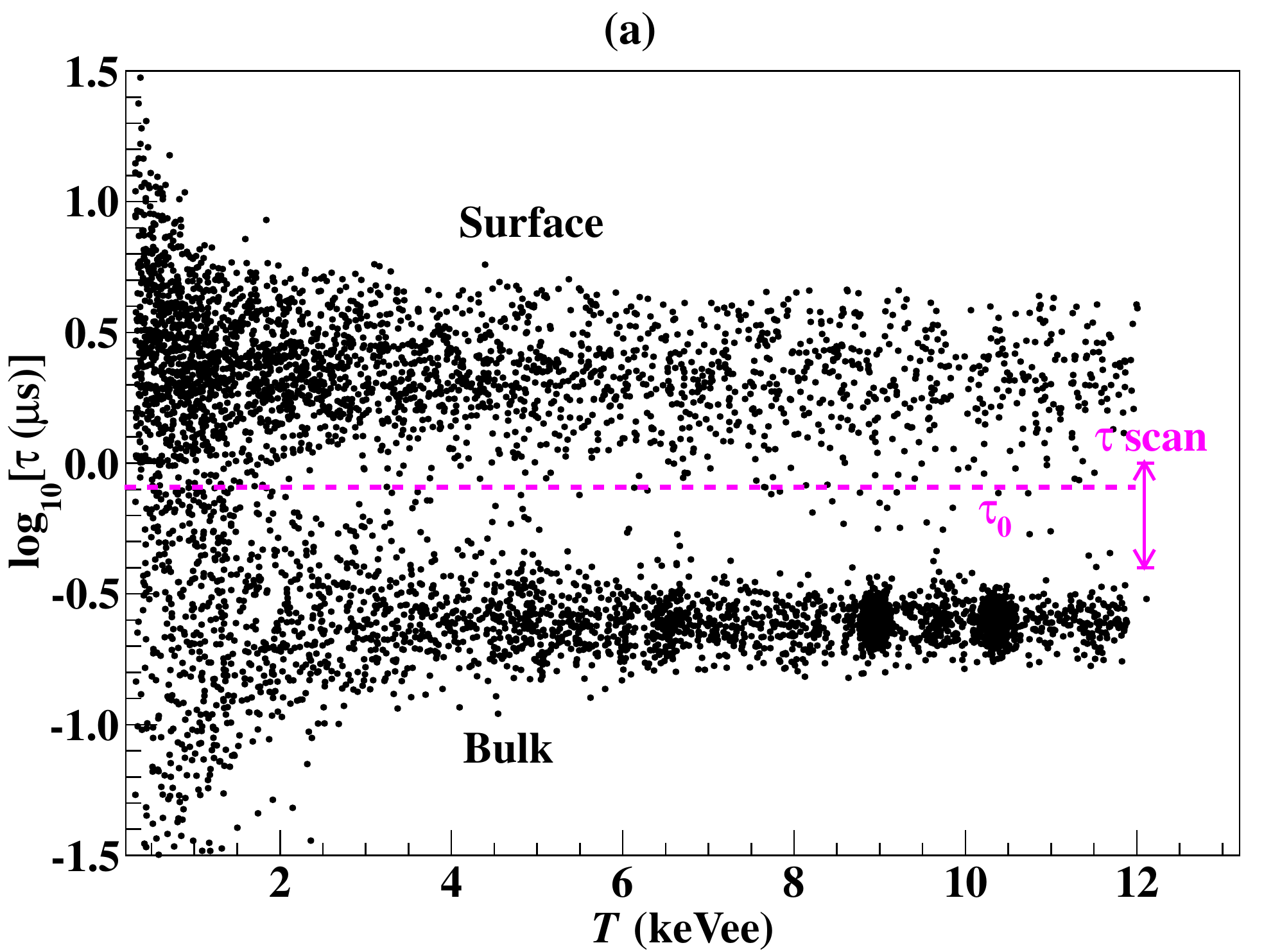}
\includegraphics[width=8.0cm,height=5.0cm]{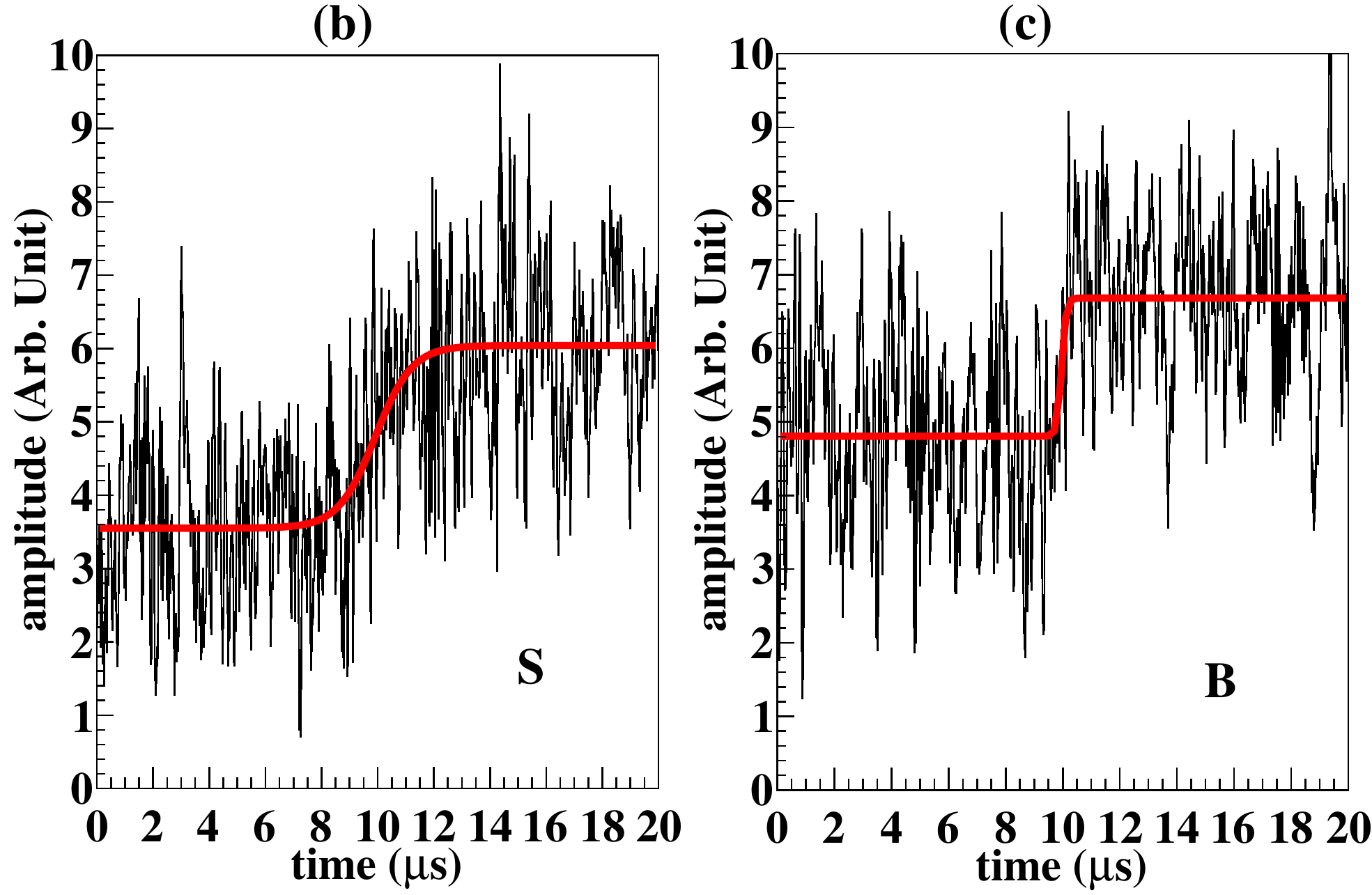}
\caption{\label{fig14bs} (a) Scatter plot of $\log_{10}(\tau)$ versus energy for AC$^{-}$ events, the BS cut criteria was defined by the $\tau_{0}$ line; (b) (c) Typical S and B events with energy at ~500 eVee, together with their fitting profiles.}
\end{figure}

We developed one data selection procedure to determine the WIMPs induced nuclear recoil events, after the dataset calibration and data quality checking~\cite{cdex1}. The procedure contains three categories of selection criteria:
\begin{enumerate}
\item Basic Cuts (BC): This basic criteria were aimed to differentiate physical events from electronic noise and spurious signals, such as microphonics. Several methods were applied to eliminate noise events according to their characteristics. The first method was based on timing information of events, derived from the distribution of parameters T$_{-}$ and T$_{+}$ , and the class of ¡°mid-period noise¡± with obvious timing distinction was identified and wiped out by the TT cut as shown in Figure~\ref{figtt}. The second method was deduced from pedestal of S$_{\textrm{p}6,12}$ and T$_{\textrm{p}}$ (Ped), which was irrelevant to the pulse shape, and therefore the criteria was defined by RT events. This method was used to discriminate the noise events whose pedestals behaved anomalously, which was mostly originated from IHB signals, as illustrated in Figure~\ref{fig11ped}. Both TT and Ped cuts are independent on event energy. The third method was dependent on pulse shape discrimination (PSD), which was on the basis of correlations of A$_{\textrm{min}}$, A$_{\textrm{max}}$, Q and T$_{\textrm{max}}$, since physical events performs different distributions in these parameters from those of noise events. The criteria was determined by physical events defined by AC$^{+}$ events of $^{137}$Cs calibration data, as depicted in Figure~\ref{fig12bc}.
\item  AC$^{+}$ versus AC$^{-}$ events selection (AC cut): Considering the $\chi$N interaction cross section, WIMPs can hardly induce signals in both $\textsl{p}$PCGe and NaI-AC detectors. However, $\gamma$ ray can produce signals in both detectors. The distribution of $\Delta$t was presented in Figure~\ref{fig13ac}. The AC$^{+}$ events with coincidence of $\textsl{p}$PCGe and NaI-AC distributed in the specific band, while AC$^{-}$ and RT events has a fixed $\Delta$t except for events with accidental coincidence. The accidental coincidence events are uniformly distributed in the time range. The trigger timing is defined by a constant amplitude discriminator of the S$_{\textrm{p6}}$ signal, such that $\Delta$t between the two detectors varies with energy.
\item Bulk versus Surface events selection (BS cut): This selection criteria was the final cut to identify AC$^{-}$ physical events which took place in the active bulk volume based on $\tau$ defined in Section IV-A. The scatter plot of $\tau$ versus energy was shown in Figure~\ref{fig14bs}(a), which emerges two characteristic bands representing bulk (B) and surface (S) events respectively. Typical B and S events as well as their fitting profile at analysis energy threshold ($\sim$500 eVee) are depicted in Figure~\ref{fig14bs}(b).
\end{enumerate}

\subsection{C. Efficiency Evaluation}
\label{4.3 efficiency}
At a total DAQ rate of $\sim$3 Hz, the DAQ live time was 99.9$\%$ measured by the survival probabilities of the RT events generated by a pulse generator at high precision and stability. Different methods have been adopted to calibrate the efficiencies for different data selection criteria. The signal efficiencies for TT, Ped and AC cuts, which are energy independent, can be evaluated by RT events accurately, and were 94.0$\%$, 96.8$\%$ and nearly 100$\%$ respectively.

The efficiency for the energy-dependent PSD cuts was derived from the physics events due to radioactive sources. Exact cuts were applied to these samples and the survival fractions provided measurements of $\varepsilon_\textrm{PSD}$ , as displayed in Figure 15a.

The final efficiency calibration is for BS cut, which required the evaluation of the B-signal retaining ($\effbs$) and S-background rejection ($\lmbdbs$) efficiencies. These two efficiency factors can translate the measured spectra (B,~S) to the actual spectra (B$_{0}$,~S$_{0}$), and their relationship was illustrated by the following coupled equations:
\begin{eqnarray}
\label{eq::elcoupled}
{\rm B} & = &  \effbs \cdot {\rm B}_0 ~  +  ~ ( 1 - \lmbdbs ) \cdot {\rm S}_0 \\
{\rm S} & = &  ( 1 - \effbs) \cdot {\rm B}_0 ~  +  ~ \lmbdbs  \cdot {\rm S}_0 ~. \nonumber
\end{eqnarray}
Since $>$99$\%$ of background from external radioactivity measured by our $\textsl{p}$PCGe detector are with energy of less than 1.5 MeV~\cite{cdex1}, $\gamma$ sources of corresponding energies [ $^{241}$Am (59.5 keV), $^{57}$Co (122keV), $^{137}$Cs (662 keV) and $^{60}$Co (1173 keV, 1332 keV) ] were used to calibrate the ($\effbs$,~$\lmbdbs$) and the detailed procedures were described in our previous work~\cite{texono2013,cdex12014,texonobs}. The energy-dependent $\effbs$ was shown in Figure 15a and $\lmbdbs$ in Figure 15b. The ($\effbs$,~$\lmbdbs$)-corrected spectra B$_0$ can be derived via Eq. (1):
\begin{eqnarray}
{\rm B}_0 & = &  \frac{\lmbdbs}{\effbs + \lmbdbs - 1} \cdot {\rm B} ~  -  ~\frac{1 - \lmbdbs}{\effbs + \lmbdbs - 1} \cdot {\rm S} ~~\\
{\rm S}_0 & = &  \frac{\effbs }{\effbs + \lmbdbs - 1} \cdot {\rm S} ~  -  ~\frac{1 - \effbs}{\effbs + \lmbdbs - 1} \cdot {\rm B} ~. \nonumber
\label{eq::elsol}
\end{eqnarray}
It was demonstrated that if the neglected (that is, taking $\lmbdbs$=1) or under-estimated S-contaminations to the B-samples can result in incorrectly-assigned signal events.

\begin{figure}[t] 
\centering
\includegraphics[height=11.0cm,width=8.0cm]{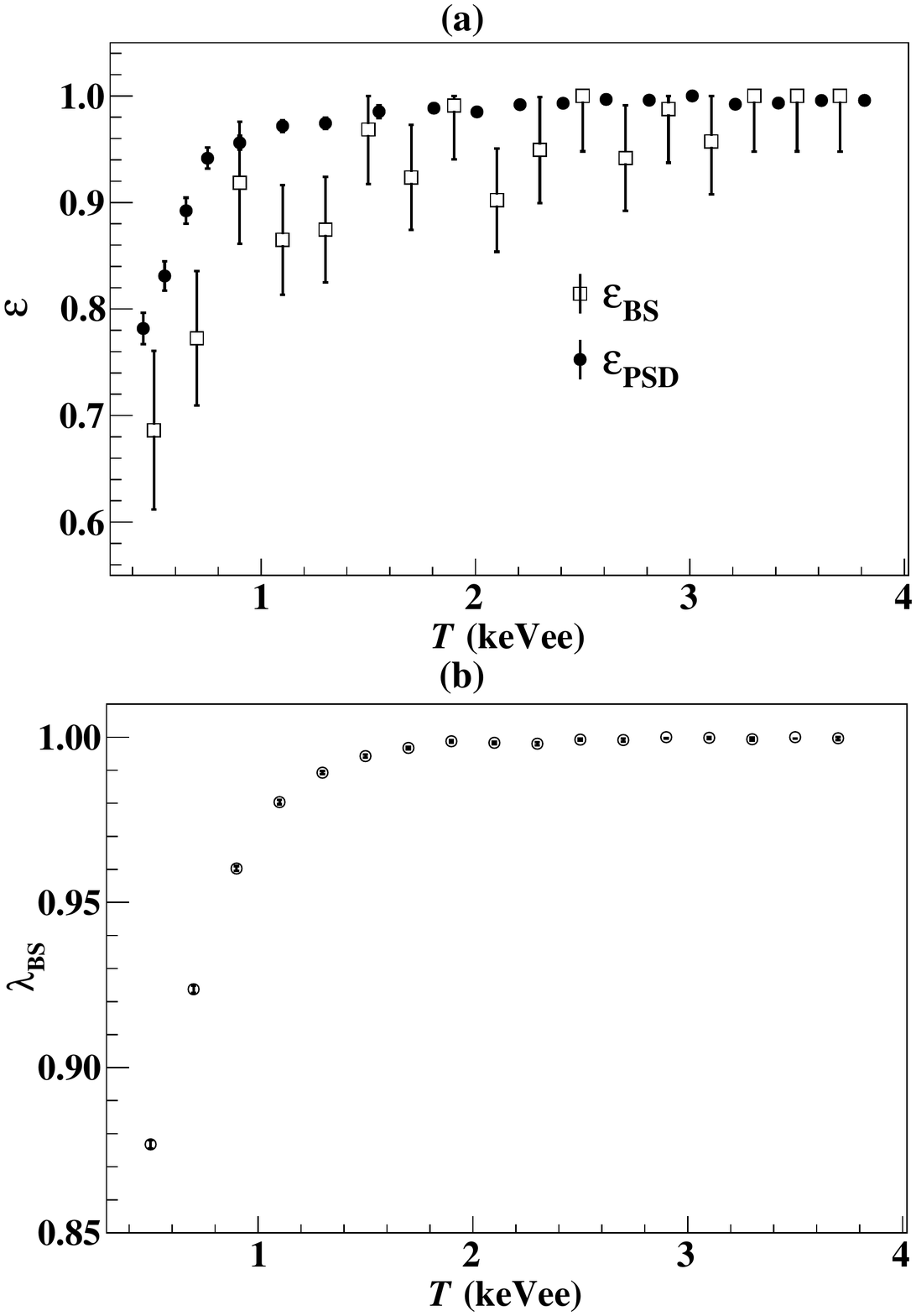}
\caption{\label{fig15eff} (a) The measured $\varepsilon_{\textrm{PSD}}$ and $\effbs$ as function of energy. (b) The measured $\lmbdbs$ as function of energy. }
\end{figure}

\subsection{D. Systematic Uncertainties}
\label{4.4 sys. error}
The systematic uncertainties of AC$^{-}$$\otimes$B$_0$ derived from raw data are summarized in Table~\ref{sys. err}, using two typical energy ranges as illustration. The systematic contributions arise from:
\begin{enumerate}
  \item Data Taking:
  \begin{enumerate}
    \item The DAQ was in stable operation at more than 98$\%$ of the time. The trigger rate is low and the DAQ live time is close to 100$\%$. Contributions to systematic uncertainties are negligible.
    \item Trigger Efficiency $-$ Since the analysis threshold (475 eVee) is much higher than the trigger threshold (246 eVee at 50$\%$), the trigger efficiency of the physics events relevant to this analysis was 100$\%$, resulting in negligible contribution to the systematic uncertainties.
    \item Fiducial Mass $-$ The error of the measured thickness of the dead layer gave rise to a 1$\%$ uncertainty at fiducial mass. This corresponds to an additional 0.1$\%$ contribution to the total systematic uncertainty at 475 eVee.
  \end{enumerate}
  \item Signal Selection: The systematic uncertainties originated from the stability of BC and AC cuts, and were studied with the change of cut parameters around the nominal values. The BC cut contributed an additional 0.5$\%$ to contribution to the total systematic uncertainty at 475 eVee, while the contribution arising from AC cut is negligible.
  \item Bulk Events Selection: The evaluation of systematic effects follow the procedures described in our earlier work~\cite{texonobs,cdex12014}.  In particular,
  \begin{enumerate}
    \item The leading systematic uncertainties is from the B-event selection and (${\rm \varepsilon_{BS}}$,~${\rm \lambda_{BS}}$) calibration due to possible differences in locations and energy spectra between the calibration sources and background events. The calibration sources probe the surface effects due to both low energy (surface richer) and high energy (bulk richer) photons. The $\tau$ distributions for B-events are identical for both sources and physics background, while those for S-events showed intrinsic difference due to the difference in surface penetration which manifest as the difference of slopes in the (${\rm \varepsilon_{BS}}$, ~${\rm \lambda_{BS}}$) plane~\cite{cdex12014}.The systematic uncertainties are derived from the spread of the (${\rm \varepsilon_{BS}}$, ~${\rm \lambda_{BS}}$) intersections of calibration bands, relative to the combined best-fit solution. This leads to a 25.0$\%$ contribution to the total error in the efficiencies-corrected Bulk rates B$_{0}$, accounting for the most significant in the total systematic uncertainty.
    \item The systematic uncertainties related with different locations are studied with the sources placed at several positions of the top and the side (the cylindrical surface) of the $\textsl{p}$PCGe. Among them, the $^{241}$Am $\gamma$ from the side are strongly attenuated due to additional thickness from the cylindrical copper support structure and curved surface of the germanium crystal and therefore do not produce useful signals. The higher energy $\gamma$ from $^{57}$Co, $^{137}$Cs and $^{60}$Co at top and side, as well as those from physics samples (BC$\otimes$AC$^{+}$ and BC$\otimes$AC$^{-}$), show similar distributions in $\tau$, independent of locations. The shift in (${\rm \varepsilon_{BS}}$,~${\rm \lambda_{BS}}$) based on calibration source data at different locations is less than 4$\%$, corresponding to a 3.7$\%$ contribution to the total error in the efficiencies-corrected Bulk rates B$_{0}$.
  \end{enumerate}
  \item Choice Quenching Function: Two studies were performed to investigate the sensitivities to exclusion limits from the choice of QF. (i) As displayed in Figure~\ref{fig7}, the red line evaluated by TRIM software together with the yellow band (10$\%$ systematic uncertainty) were adopted. Analysis is performed by scanning QF within 10$\%$ of their nominal value. It was shown that the difference among these results are small, eg. the variation of $\sigma^{SI}_{\chi N}$ is about 15$\%$ at $m_{\chi}$=8 GeV/c$^{2}$, and the least stringent bounds among them at a given WIMP-mass were adopted as our final physics limits. (ii) The same procedure described in (i) but the QF evaluated by Lindhard ($\textit{k}$=0.157) and CoGeNT~\cite{cogent} were applied. It was concluded that the difference were small, eg. about 14$\%$ deviations in $\sigma^{SI}_{\chi N}$ at m$_{\chi}$=8 GeV/c$^{2}$). The results have been displayed in~\cite{cdex12014} and our formulation with TRIM provided the most conservative limits among the alternatives.
\end{enumerate}

In our previous work, the 53.9 kg-days exposure has shown that the statistical uncertainties were dominant and contributed 86$\%$ relative to the total uncertainty~\cite{cdex12014}.  As the exposure expanded to 335.6 kg-days, the statistical uncertainties were secondary and systematic uncertainties dominated 81$\%$ relative to the total uncertainty. It was crucial to develop new method to evaluate the (${\rm \varepsilon_{BS}}$, ${\rm \lambda_{BS}}$) to decrease the systematic uncertainty further which contributes the main part of the systematic uncertainty.

\renewcommand\tablename{TABLE}
\renewcommand{\thetable}{\arabic{table}}
\begin{table*}
\begin{ruledtabular}
\caption{\label{sys. err} The various contributions to the total error of AC$^-$$\otimes$B$_0$
at threshold and at a typical high energy bin.}
\centering
\begin{tabular}{lcc}
Energy Bin & 0.475-0.575 keVee & 1.975-2.075 keVee \\
AC$^-$$\otimes$B$_0$ and Errors & $ 4.00 \pm 0.64 [$\rm stat$] \pm 0.87 [$\rm sys$]$ & $3.61 \pm 0.36 [$\rm stat$] \pm 0.28 [$\rm sys$]$ \\
(kg$^{-1}$keV$^{-1}$day$^{-1}$) & $=4.00 \pm 1.08 $ & $=3.61 \pm 0.46 $ \\
\hline
I) Statistical Uncertainties : \\
    \hspace{0.5cm}(i)Uncertainties on Calibration (${\rm \varepsilon_{BS}}$,${\rm \lambda_{BS}}$) : & \hspace{0cm}0.32 & \hspace{0cm}0.08 \\
    \hspace{0.5cm}(ii)Derivation of (${\rm \varepsilon_{BS}}$,${\rm \lambda_{BS}}$)-corrected Bulk Rates : & \hspace{0cm}0.55 & \hspace{0cm}0.35 \\
    \hspace{0.5cm}Combined : & 0.64 & 0.36 \\
\hline
II) Systematic Uncertainties : \\
    \hspace{0.5cm}A. Data Taking : \\
    \hspace{0.8cm}(i) DAQ : & 0.00 & 0.00 \\
    \hspace{0.8cm}(ii) Trigger Efficiency : & 0.00 & 0.00 \\
    \hspace{0.8cm}(iii) Fiducial Mass : & \hspace{0cm}0.05 & \hspace{0cm}0.05 \\
    \hspace{0.5cm}B. Signal Selection : \\
    \hspace{0.8cm}(i) BC cuts : & 0.08 & 0.05 \\
    \hspace{0.8cm}(ii) AC cut : & 0.00 & 0.00 \\
    \hspace{0.5cm}C. Bulk Event Selection : \\
    \hspace{0.8cm}(i) Rise-time Cut-Value $\tau_{0}$ & \hspace{0cm}0.27 & \hspace{0cm}0.12 \\
    \hspace{0.8cm}(ii) Normalization Range (3-5 keVee) & \hspace{0cm}0.07 & \hspace{0cm}0.01 \\
    \hspace{0.8cm}(iii) (${\rm B_{0}}$,${\rm S_{0}}$) = (B,S) at Normalization & \hspace{0cm}0.10 & \hspace{0cm}0.10 \\
    \hspace{0.8cm}(iv) Choice of Discard Region & \hspace{0cm}0.30 & \hspace{0cm}0.06 \\
    \hspace{0.8cm}(v) Source Location & \hspace{0cm}0.28 & \hspace{0cm}0.19 \\
    \hspace{0.8cm}(vi) Source Energy Range and Spectra & \hspace{0cm}0.72 & \hspace{0cm}0.12 \\
    \hspace{0.5cm}Combined : & 0.87 & 0.28 \\
\end{tabular}
\end{ruledtabular}
\end{table*}

\section{V. Limits on WIMPs}
\label{5. exclusion physical results}
\begin{figure}[t] 
\centering
\includegraphics[width=8.0cm,height=10.0cm]{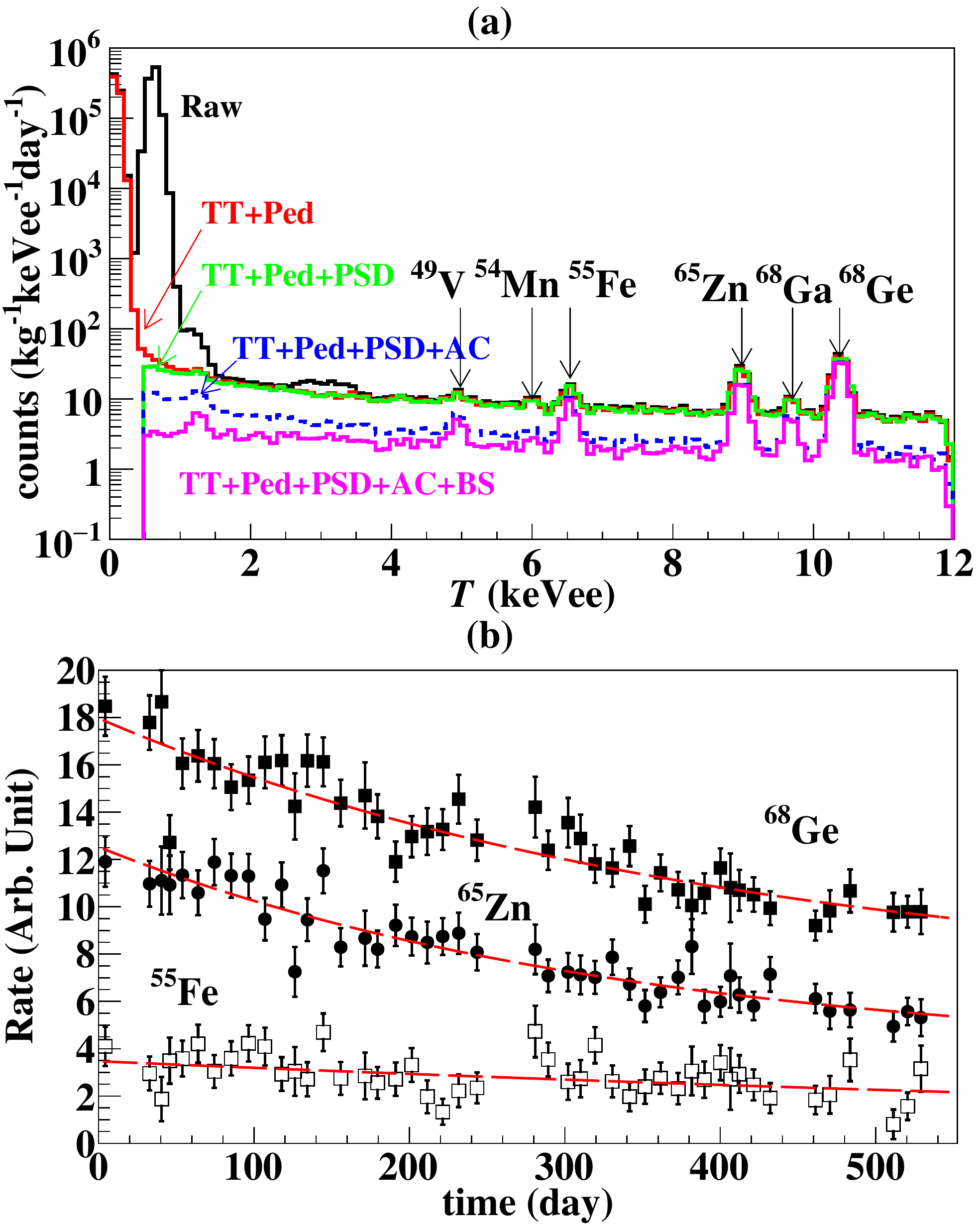}
\caption{\label{fig16sp} (a) Measured energy spectra of 1kg-$\textsl{p}$PCGe detector and its evolution with data selection progress. Six cosmogenic nuclides have been identified. (b) Time evolution of the three dominant K-shell X-rays: $^{68}$Ge, $^{65}$Zn and $^{55}$Fe. The measured half-lifes 279.7$\pm$17.8 days, 235.3$\pm$16.0 days and 955.5$\pm$411.2 days, respectively, are consistent with the nominal values of 270.8 days, 244.3 days and 997.1 days. }
\end{figure}

\begin{figure}[t] 
\includegraphics[width=7.0cm,height=13.0cm]{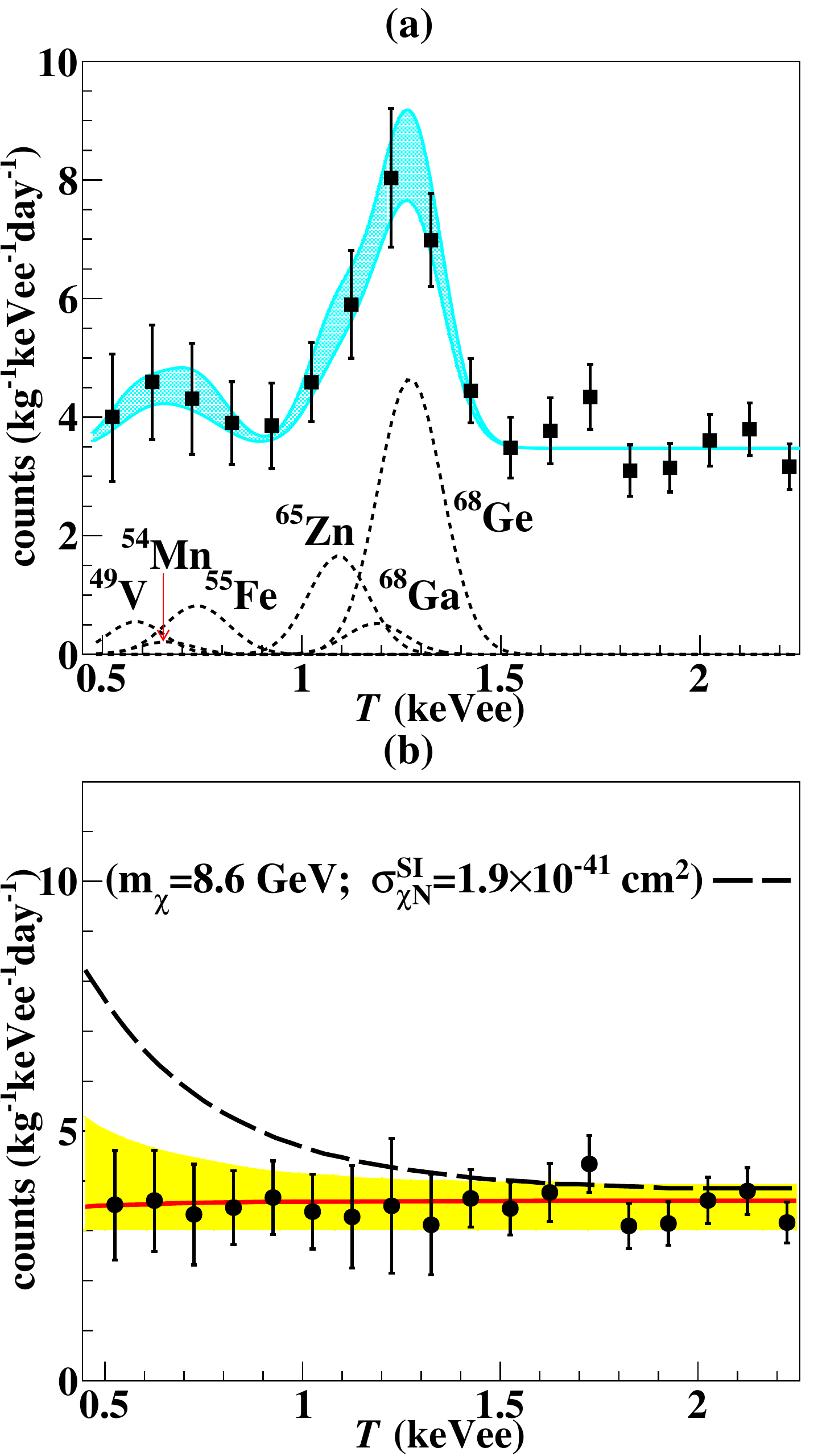}
\caption{\label{fig17lxres} (a) Energy spectrum with all selection cuts and efficiency correction factors applied. Various L-shell X-rays are identified based on measured K-shell X-rays intensities, and superimposed on a flat background from ambient high-energy gamma-rays. (b) The residual spectrum with contributions subtracted.The red-line represent the best-fit with two parameters: flat gamma-background and spin-independent $\chi$-N cross-section, at m$_{\chi}$=8 GeV/c$^{2}$. An excluded (m$_{\chi}$;~$\sigma^{SI}_{\chi N}$) scenario of CDMS(Si)~\cite{cdms2si} is superimposed. }
\end{figure}

The measured energy spectra and its evolution with data selection progress are depicted in Figure~\ref{fig16sp} (a).  Six cosmogenic nuclides can be identified clearly through the K-shell X-rays peaks, and the contributions of the corresponding L-shell X-rays at low energy range can be calculated accurately since the ratioes of the intensities of K-shell and L-shell X-rays are definite, as shown in Figure~\ref{fig17lxres} (a).  The half-lives of the dominant nuclides can be measured by their K-shell X-rays rays. Figure~\ref{fig16sp} (b) displayed the decay of $^{68}$Ge, $^{65}$Zn and $^{55}$Fe. High energy $\gamma$ rays originated from ambient radioactivity contributed a flat electron-recoil background.

The nature of the interaction between WIMPs with baryonic matter is a priori unknown. The data was analysed with two benchmark $\chi$-N cross-sections: spin-independent (SI, scalar) and spin-dependent (SD, axial-vector) couplings:
\begin{eqnarray}
\label{eq::SISD}
{\rm \frac{d\sigma_{\chi N}}{dE_{R}}} & = &  {\rm (\frac{d\sigma_{\chi N}}{dE_{R}})_{SI} ~  +  ~(\frac{d\sigma_{\chi N}}{dE_{R}})_{SD}} ~.
\end{eqnarray}
In general, the SI cross section can be written as:
\begin{eqnarray}
\label{eq::SI}
{\rm (\frac{d\sigma_{\chi N}}{dE_{R}})_{SI}} & = &  {\rm \frac{2 m_{N}}{\pi \emph{v}^{2}}[Z\emph{f}_{p} ~ + ~ (A ~ - ~Z)\emph{f}_{n}]^{2}F^{2}(E_{R}) } ~.
\end{eqnarray}
where the $\emph{f}_{\textrm{p,n}}$ describe the WIMPs couplings to proton and neutron. In most cases $\emph{f}_{\textrm{p}}$  $\approx$  $\emph{f}_{\textrm{n}}$, and the Eq. (4) can be simplified to:
\begin{eqnarray}
\label{eq::SIsi}
{\rm (\frac{d\sigma_{\chi N}}{dE_{R}})_{SI}} & = &  {\rm \frac{2 m_{N}}{\pi \emph{v}^{2}} A^{2} (\emph{f}_{p})^{2} F^{2}(E_{R}) } ~.
\end{eqnarray}
leading to $\textrm{A}^{2}$ dependence of the SI cross section. The SD differential cross section can be expressed as:
\begin{eqnarray}
\label{eq::SD}
{\rm (\frac{d\sigma_{\chi N}}{dE_{R}})_{SD}} & = &  {\rm \frac{16 m_{N}}{\pi \emph{v}^{2}} \Lambda^{2} G_{F}^{2} J(J  +  1) \frac{S(E_{R})}{S(0)} }~.
\end{eqnarray}
where the J is the total angular momentum of the nucleus. The Eq. (6) illustrates that the SD cross section is proportional to  a function of the total angular momentum of the nucleus, J/(J+1)~\cite{sisd}.

\begin{figure}[t] 
{\bf (a)}\\
\includegraphics[width=8.0cm,height=8.0cm]{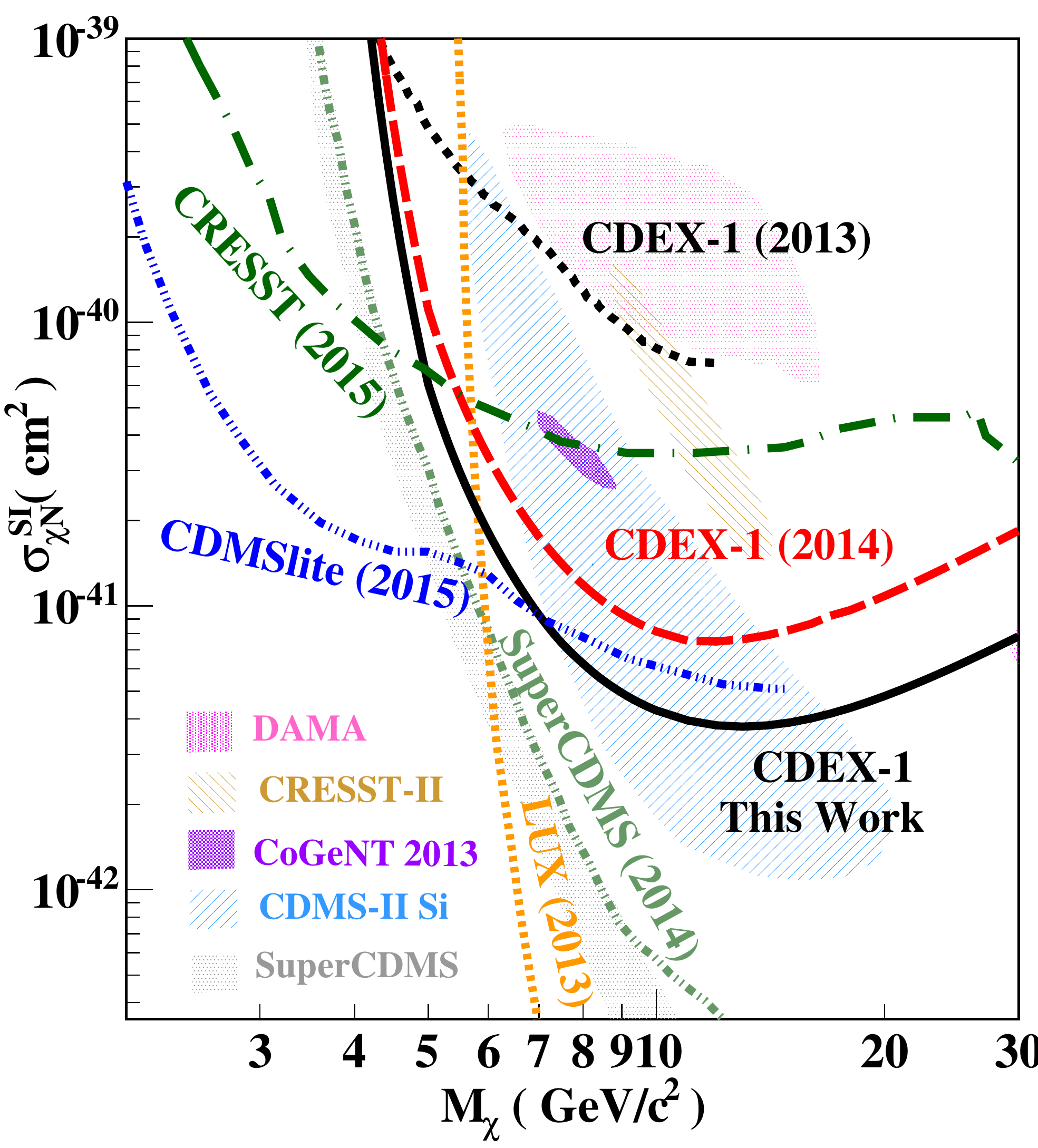}\\
{\bf (b)}\\
\includegraphics[width=8.0cm,height=8.0cm]{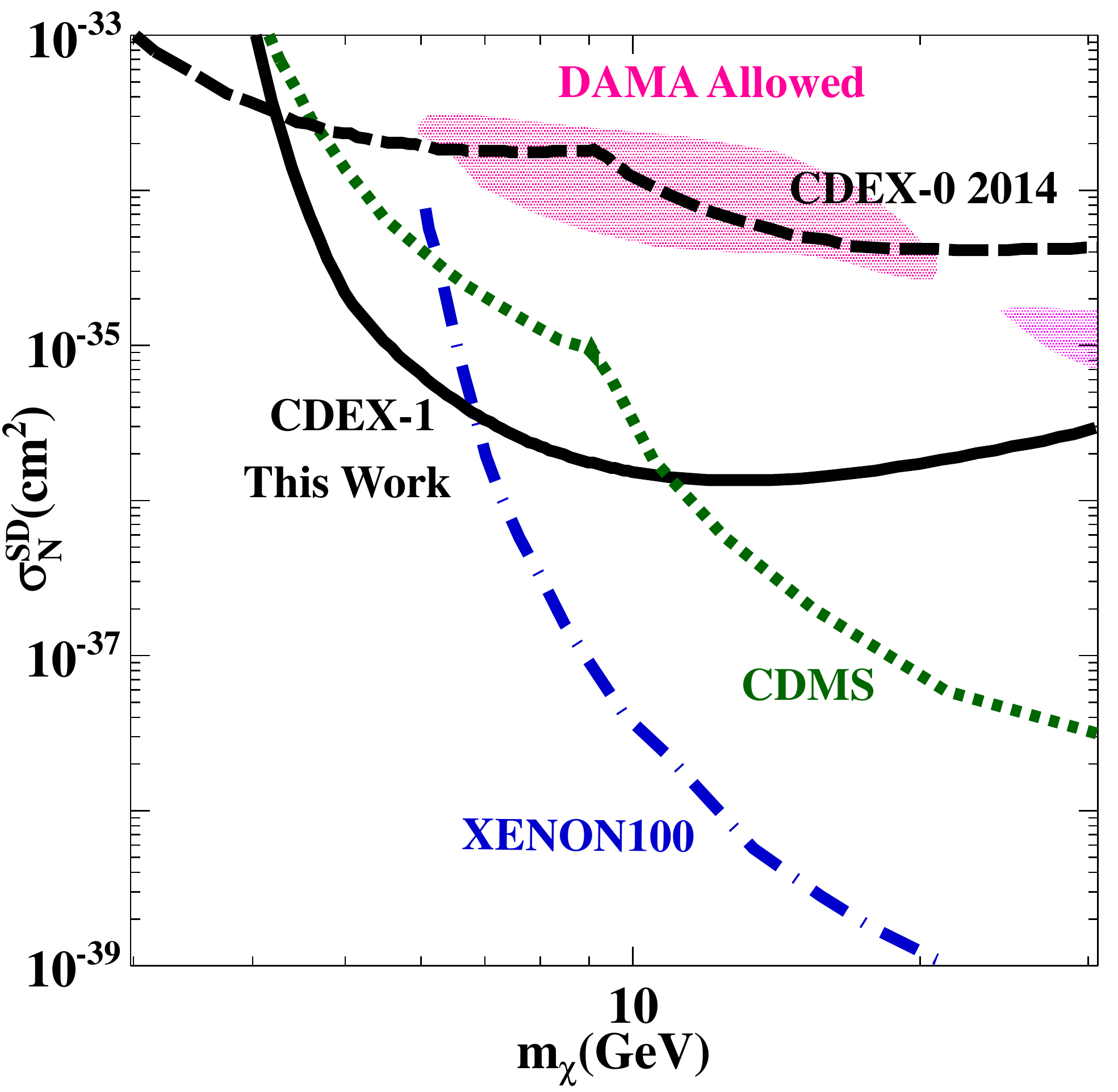}
\caption{ The 90$\%$ confidence level upper limit of (a) spin-independent $\chi$-N coupling and (b) spin-dependent $\chi$-neutron cross-sections. The CDEX-1 results from this work are depicted in solid black. Bounds from other benchmark experiments~\cite{cdex1,cdex12014,cogent,dama,cdmslite,lux,supercdms,cresst2015,cdms2si} are superimposed.
}
\label{fig18exp}
\end{figure}

A best-fit analysis was applied to the residual spectrum of Figure 17b after subtraction of the L-shell X-rays, with two parameters representing flat gamma background and possible $\chi$N spin-independent cross section $\sigma^{SI}_{\chi N}$, scanning m$_{\chi}$ between 4 and 30 GeV/c$^{2}$. Standard WIMP halo assumption~\cite{wimphalo} and conventional astrophysical models~\cite{wimpmodel} are applied to describe WIMP-induced interactions, with the local WIMP density of 0.3 GeV/cm$^{3}$, the Maxwellian velocity distribution with $v_{0}$=220 km/s and the galactic escape velocity of $v_{esc}$=544 km/s. Exclusion plots on (m$_{\chi}$,~$\sigma^{SI}_{\chi N}$) at 90$\%$ confidence level were shown in Figure~\ref{fig18exp}(a), together with bounds and allowed regions from several representative experiments~\cite{cogent,dama,cdmslite,lux,supercdms,cresst2015,cdms2si}. The sensitivities of $\sigma^{SI}_{\chi N}$ has been improved a few times over our work last year~\cite{cdex12014} due to several times larger exposure. Most of the light WIMP regions within 6 and 20 GeV/c$^2$ implied by earlier experiments are probed and rejected.

The limits on spin-dependent $\chi$-neutron (denoted by $\chi$n) cross sections were also extracted. Exclusion plots on (m$_{\chi}$,$\sigma^{SD}_{\chi n}$) plane at 90$\%$ confidence level for light WIMPs was also derived, as depicted in Figure~\ref{fig18exp}(b), and bounds from other benchmark experiments~\cite{dama_sd_allowed,cdms_sd_le,xenon100_sd} are also superimposed. The limits were derived from the model-independent approaches prescribed in Refs~\cite{sd1,sd2}. Different $^{73}$Ge nuclear physics matrix elements~\cite{geme} adopted as input generated consistent results. The DAMA allowed region at low-m$_{\chi}$ was probed and excluded. Furthermore, it was shown that these results were competitive around m$_{\chi}$=6 GeV/c$^{2}$. For completeness, the exclusion limits for the spin-dependent cross-section derived from our earlier CDEX-0 data~\cite{cdex0} are also displayed in Figure~\ref{fig18exp}(b).

\section{VI. Summary and Prospects}
\label{6. conclusion}
The hardware, operation and analysis details of the CDEX-1 experiment are described in this article. New limits on both SI and SD cross-sections are derived with a data size of 335.6 kg-days, spanning over 17 months. The studies of annual modulation effects with this data set are being pursued. Another 1 kg $\textsl{p}$PCGe with lower threshold is taking data at CJPL with data analysis and background understanding underway.

A $\textsl{p}$PCGe ``CDEX-10'' detector array with target mass of the range of 10 kg and installed in liquid nitrogen as cryogenic medium is being commissioned. A future option of replacement with liquid argon to serve in addition an anti-Compton detector is being explored. In the meantime, a $\textsl{p}$PCGe detector completely fabricated by the CDEX Collaboration with a Ge crystal provided by the Industry is being constructed. This would allow complete control on the choice of materials which are crucial towards the future goal of ton-scale Ge detectors for dark matter and double beta decay experiments.

\section{Acknowledgements}
\label{7. thanks}
This work was supported by the National Natural Science Foundation of China (Contracts No.10935005, No.10945002, No.11275107, No.11175099, No.11475099) and National Basic Research program of China (973 Program) (Contract No. 2010CB833006) and NSC 99-2112-M-001-017-MY3 and Academia Sinica Principal Investigator 2011-2015 Grant from Taiwan.


\begin{thebibliography}{99}
\bibitem{cdextarget}
K.J. Kang et al., Front. Phys. {\bf 8}, 412 (2013).
\bibitem{cjpl}
K. J. Kang, J. P. Cheng, Y. H. Chen, Y. J. Li, M. B. Shen, S. Y. Wu, and Q. Yue, J. Phys. Conf. Ser. {\bf 203}, 012028 (2010).
\bibitem{rpp}
K.A.Olive et al., Review of Particle Physics, Chin. Phys. C, {\bf 38} 090001 (2014).
\bibitem{dm2015}
Marc Schumann., EPJ Web of Conferences {\bf 96} 01027 (2015).
\bibitem{dm_dbd}
P. Cushman, C. Galbiati, et al., arXiv:1310.8327.
\bibitem{dbd2015}
A.S. Barabash., Physics Procedia {\bf 74} (2015) 416šC422.
A.S. Barabash., AIP Conf.Proc. {\bf 1686} (2015) 020003.
\bibitem{dbd2012}
S.M. Bilenky, C. Giunti., Mod. Phys. Lett. A {\bf 27}, 1230015 (2012).
S.R. Elliott, Mod. Phys. Lett. A, {\bf 27}, 1230009 (2012).
\bibitem{cdex0}
S. K. Liu et al., Phys. Rev. D {\bf 90}, 032003 (2014).
\bibitem{cdex1}
W. Zhao et al., Phys. Rev. D {\bf 88}, 052004 (2013);
K. J. Kang et al., Chin. Phys. C {\bf 37}, 126002 (2013).
\bibitem{texono2003-2007}
H.B. Li et al, PRL {\bf 90}, 131802 (2003);
H.T. Wong et al., PRD {\bf 75}, 012001 (2007).
\bibitem{cdex12014}
Q. Yue et al., Phys. Rev. D {\bf 90}, 091701(R) (2014).
\bibitem{cogent}
C. E. Aalseth et al., Phys. Rev. D {\bf 88}, 012002 (2013); arXiv:1401.3295.
\bibitem{cjplmuon}
Y. C. Wu et al., Chin. Phys. C {\bf 37}, 086001 (2013).
\bibitem{cjplgamma}
Zhi Zeng et al., J Radioanal Nucl Chem {\bf 301}:443-450(2014).
\bibitem{texono2013}
H. B. Li et al., Phys. Rev. Lett. {\bf 110}, 261301 (2013).
\bibitem{trim}
J. F. Ziegler, http://www.srim.org.
\bibitem{lindard}
J. Lindhard et al., K. Dan. Vidensk. Selsk. Mat. Fys. Medd. {\bf 33}, 10 (1963).
\bibitem{texonobs}
H.B. Li et al. Artropart. Phys. {\bf 56}, 1 (2014).
\bibitem{deadlayer_majorana}
E. Aguayo et al., Nucl. Inst. Meth. A {\bf 701}, (2013).
\bibitem{texono2009}
S. T. Lin et al., Phys. Rev. D {\bf 79}, 061101 (2009).
\bibitem{cdmslite}
R. Agnese et al., arXiv:1509.02448v1.
\bibitem{sisd}
David G. Cerdeno and Anne M. Green., PARTICLE DARK MATTER:Observations, Models and Searches. {\bf 17}:347-352 (2010).
\bibitem{wimphalo}
F. Donato, N. Fornengo, and S. Scopel, Astropart. Phys. {\bf 9},247 (1998).
\bibitem{wimpmodel}
M. Drees and G. Gerbier, Phys. Rev. D {\bf 88}, 012002 (2013), and references therein.
\bibitem{dama}
R.~Bernabei et al., Eur. Phys. J. C {\bf 56}, 333 (2008);
R.~Bernabei et al., Eur. Phys. J. C {\bf 67}, 39 (2010).
\bibitem{cresst2015}
G. Angloher et al., Eur. Phys. J. C {\bf 72}, 1971 (2012).
G. Angloher, et al., arXiv:1509.01515v1.
\bibitem{cdms2si}
R. Agnese et al., Phys. Rev. Lett. {\bf 111}, 251301 (2013).
\bibitem{lux}
D. S. Akerib et al., Phys. Rev. Lett. {\bf 112}, 091303 (2014).
\bibitem{supercdms}
R. Agnese et al., Phys. Rev. Lett. {\bf 112}, 241302 (2014).
\bibitem{dama_sd_allowed}
C. Savage et al., arXiv:0808.3607v2.
\bibitem{cdms_sd_le}
Z. Ahmed et al., Phys. Rev. Lett. {\bf 106}, 131302 (2011), arXiv:1011.2482v3.
Z. Ahmed et al., Phys. Rev. Lett. {\bf 102}, 011301 (2009), arXiv:0802.3530v2.
\bibitem{xenon100_sd}
E. Aprile et al., arXiv:1301.6620v2.
\bibitem{sd1}
A. Bottino et al., Phys. Lett. B {\bf 402}, 113 (1997).
\bibitem{sd2}
D. R. Tovey et al., Phys. Lett. B {\bf 488}, 17 (2000).
\bibitem{geme}
M. T. Ressell et al., Phys. Rev. D {\bf 48}, 5519 (1993); V. I. Dimitrov, J. Engel, and S. Pittel, Phys. Rev. D {\bf 51}, R291 (1995).


\end{thebibliography}
\end{document}